\documentclass[aps,pra,twocolumn,showpacs]{revtex4-1}

\usepackage[dvips]{graphicx}
\usepackage{caption}
\usepackage{subcaption}

\usepackage{amsmath,amssymb}
\usepackage{amsfonts}
\usepackage{color}

\usepackage[normalem]{ulem} 
\usepackage{placeins}
\usepackage{mdframed} 
\usepackage{soul} 
\usepackage{comment} 

\usepackage{tikz}
\usepackage{pgfplots}
\usepackage{cleveref}
\usepackage{float}
\usepackage{soul}
\usepackage{lipsum}
\usepackage{readarray}

\newcommand\eg{\textit{e.g.}}
\newcommand\ie{\textit{i.e.}}
\newcommand\etal{\textit{et al.}}

\newcommand\e{\ensuremath{\mathrm{e}}}
\newcommand\I{\ensuremath{\mathrm{i}}}

\renewcommand\Im{\ensuremath{\text{Im}}}
\newcommand\rd{\ensuremath{\mathrm{d}}}

\newcommand\J[1][]{\ensuremath{_{\mathrm{J}#1}}}
\newcommand\dc{\ensuremath{_\mathrm{dc}}}

\newcommand\rf{\ensuremath{_\mathrm{rf}}}
\newcommand\crit[1][]{\ensuremath{_{\mathrm{c}#1}}}

\newcommand\pump{\ensuremath{_\mathrm{p}}} 
\newcommand\ph[1]{\ensuremath{_{#1\mathrm{p}}}} 
\newcommand\pAny[1]{\ensuremath{_{\mathrm{p}#1}}} 
\newcommand\signal[1][]{\ensuremath{_{\mathrm{s}#1}}} 
\newcommand\sh[1]{\ensuremath{_{\mathrm{s}+#1\mathrm{p}}}} 
\newcommand\sAny[1]{\ensuremath{_{\mathrm{s}#1}}} 
\newcommand\idler{\ensuremath{_\mathrm{i}}} 
\newcommand\ih[1]{\ensuremath{_{\mathrm{i}+#1\mathrm{p}}}} 
\newcommand\iAny[1]{\ensuremath{_{\mathrm{i}#1}}} 

\newcommand\signed[1]{\ifnum #1>0 {+#1} \else{\ifnum #1<0 {#1} \fi} \fi}
\newcommand\coeff[1]{
    \ifnum #1=1 {+} \else{
        \ifnum #1=-1 {-} \else{
            \ifnum #1>0 {+#1} \else{
                \ifnum #1<0 {#1} \fi
            } \fi
        } \fi
    } \fi
}

\input{settings/tikzicons.tex}

\Crefname{equation}{Eq.}{Eqs.}
\Crefname{figure}{Fig.}{Figs.}
\Crefname{tabular}{Tab.}{Tabs.}
\pgfplotsset{compat=1.3}
\setstcolor{red}

\newcommand\freqscale{3}
\newcommand\mixheightlower{0.75}
\newcommand\mixheightupper{0.6}
\newcounter{ph} 

\definecolor{mycolor1}{rgb}{0.00000,0.44700,0.74100}%
\definecolor{mycolor2}{rgb}{0.85000,0.32500,0.09800}%
\definecolor{mycolor3}{rgb}{0.92900,0.69400,0.12500}%
\definecolor{mycolor4}{rgb}{0.49400,0.18400,0.55600}%
\definecolor{mycolor5}{rgb}{0.46600,0.67400,0.18800}%
\definecolor{mycolor6}{rgb}{0.30100,0.74500,0.93300}%
\definecolor{mycolor7}{rgb}{0.63500,0.07800,0.18400}%
\definecolor{BLUE}{rgb}{0, 0, 1}

\begin{document}

\title{A high gain travelling-wave parametric amplifier based on three-wave mixing}
\author{Hampus Renberg Nilsson}
\email{Hampus.Renberg.Nilsson@chalmers.se}
\author{Anita Fadavi Roudsari}
\author{Daryoush Shiri}
\author{Per Delsing}
\author{Vitaly Shumeiko}
\address{Department of Microtechnology and Nanoscience - MC2, Chalmers University of Technology,
S-412 96 G\"oteborg, Sweden.}
\date\today

\begin{abstract}
We extend the theory for a Josephson junction travelling wave parametric amplifier (TWPA) operating in the three-wave mixing regime and we propose a scheme for achieving high gain. The continuous three-mode model [P. K. Tien, J. Appl. Phys. \textbf{29}, 1347 (1958)] is on one hand extended to describe a discrete chain of Josephson junctions at high frequencies close to the spectral cutoff where there is no up-conversion. On the other hand, we also develop a continuous multimode theory for the low-frequency region where the frequency dispersion is close to linear. We find that in both cases the gain is significantly reduced compared to the prediction by the continuous three-mode model as the result of increasingly strong dispersion at the high frequencies and generation of up-converted modes at the low frequencies. The developed theory is in quantitative agreement with experimental observations. To recover the high gain, we propose to engineer a chain with dispersive features to form a two-band frequency spectrum and to place the pump frequency within the upper band close to the spectral cutoff. We prove that there exists a sweet spot, where the signal and the pump are phase matched, while the up-conversion is inhibited. This results in a high gain which grows exponentially with the length of the TWPA.
\end{abstract}

\maketitle

\section{Introduction}

Quantum limited parametric amplifiers \cite{Caves1982} are important tools for measuring and monitoring states of superconducting qubits. Built with nonlinear, superconducting lumped element oscillators or transmission line resonators and demonstrating high gain and small added noise \cite{Yamamoto2008,Bergeal2010,Roch2012,Roy2016,AumentadoRev}, the parametric amplifiers became an essential part of the circuit Quantum Electrodynamics (cQED) \cite{Girvin2008} toolbox. 

To build a large-scale multiqubit quantum processor, an optimisation of qubit readout by multiplexing is desirable, which requires amplifiers with a large bandwidth, high gain and low added noise. Such a capability is provided by travelling wave parametric amplifiers (TWPA). During the last few years, the interest has rapidly grown in the development and investigation of the properties of different types of TWPA. 

The amplification principle of the TWPA is based on nonlinear interaction of a weak propagating signal with an intense co-propagating wave (pump), which under a phase-matching condition results in an exponential spatial growth in the signal amplitude \cite{Cullen1958,Suhl1958,Tien1958}. In the quantum regime, the TWPA is capable to generate signal squeezing and photon entanglement \cite{Grimsmo2017,Aalto2021}. 

Within the cQED platform, two types of TWPAs are experimentally tested and theoretically studied: the first type uses the nonlinear kinetic inductance of a superconducting transmission line \cite{LeDuc2012,Pappas2014,Pappas2016,Gao2017,Erickson2017,Katz2020,GaoPRXQ2021, Duti2021}, and the second uses the nonlinear inductance of a chain of Josephson junctions  \cite{Siddiqi2013,Obrien2014,Macklin2015,Martinis2015,Bell2015,Planat2020,Ranadive2021,Zorin2016,Zorin2019,Sivak2019,Mukhanov2019,Aalto2021}. The TWPAs are further distinguished depending on the type of nonlinear interaction they employ: three-wave mixing (3WM) or four-wave mixing (4WM).
 
In this paper, we theoretically investigate the efficiency of the TWPA operating in the 3WM regime,  we compare and verify our theoretical model with experimental data and we suggest a way to achieve high gain.
 
The 3WM amplifiers employ the lowest order, cubic, nonlinearity of the inductive energy, which is similar to the \(\chi^{(2)}\) nonlinearity in optical crystals. Such nonlinearity is associated with the broken time-reversal symmetry, which can be introduced by applying a dc current bias, or a magnetic flux bias. The amplification occurs due to a down-conversion process, which is capable to provide an efficient amplification within a large bandwidth in a weakly dispersive medium already at relatively small pump intensity \cite{Tien1958}. An important property of this regime is the separation of the amplification band from the pump, and also the possibility of phase preserving as well as phase sensitive amplification.

In practice, however, the amplification performance of 3WM devices with weak frequency dispersion is compromised by the generation of pump harmonics \cite{Bloembergen1962} as well as signal and idler up-conversion \cite{Pappas2016,Dixon2020}. 

The 4WM amplifiers employ the next order, quartic, nonlinearity of the inductive energy, which is similar to the \(\chi^{(3)}\) nonlinearity in optical fibers. Amplification in this regime is less efficient since it is a higher order effect with respect to the pumping strength, and it also suffers from dephasing due to Kerr effect that makes exponential amplification impossible without dispersion engineering \cite{Obrien2014,Macklin2015,Martinis2015}. Furthermore, the pump position in the middle of the gain band is undesirable for certain applications.

Our quest in this paper is to investigate whether it is possible to realise in practice full exponential amplification in a TWPA using 3WM by avoiding the poisoning effect of up-converted modes.

At first glance, the discreteness of the Josephson junction chain allows solving of the problem by placing the pump frequency close to the spectral cutoff, inherent in the discrete chains, and thus eliminating the up-converted modes. Our analysis based on the exact solution for the discrete chain shows that this is in principle possible. However, to overcome the effect of dispersion, which becomes increasingly strong in the vicinity of the spectral cutoff, a rather strong pump signal is required that is unlikely to be realised in practice.

We propose to solve this difficulty by engineering a two-band frequency spectrum of the TWPA and placing the pump within the upper band close to the spectral cutoff. In this case, as we prove, there exists a sweet spot where the pump and the signal belong to the different bands and are exactly phase-matched, while the generation of up-converted modes is inhibited since the pump is close enough to the cutoff. In the vicinity of this sweet spot, a rather broad window opens where a strong exponential amplification occurs. The width of this window is limited by the dispersion and up-conversion effects (c.f. experimental observation in a kinetic inductance TWPA in \cite{GaoPRXQ2021}). 

The structure of the paper is the following. In \Cref{sec:equations} we derive universal dynamic equations for three different kinds of TWPAs, that use either current biased junctions, or magnetic flux biased radio-frequency superconducting quantum interference devices (rf-SQUIDs) \cite{Zorin2016}, or magnetic flux biased superconducting nonlinear asymmetric inductive elements (SNAILs) \cite{Sivak2019}.

In \Cref{sec:Disc3modes} we derive the exact solution to the model containing only three modes involved in the down-conversion, while neglecting up-converted modes, and we evaluate the exponential gain and the frequency region where the exponential gain exists. This part is a generalisation of the solution for a continuous medium in Ref.~\cite{Tien1958}. Here we also evaluate the pumping strength required for this model to be valid.

In \Cref{sec:quasilinear} we investigate the low frequency region with weak frequency dispersion and show that generation of up-converted modes makes it impractically hard to achieve high exponential gain (cf. \cite{Dixon2020}). We also compare our theoretical results with experimental data obtained on a SNAIL-based TWPA, and find a very good quantitative agreement. 

In \Cref{sec:solution} we pursue the strategy of boosting the gain by engineering a two-band spectrum of the TWPA. We identify the sweet spots, where the high gain is achieved, for the two TWPA designs - adding resonators to unit cells \cite{Obrien2014,Macklin2015,Martinis2015}, and periodically modulating the TWPA parameters \cite{Planat2020}. The obtained results are summarised in \Cref{sec:conclusion}.

\section{TWPA dynamical equations}
\label{sec:equations}

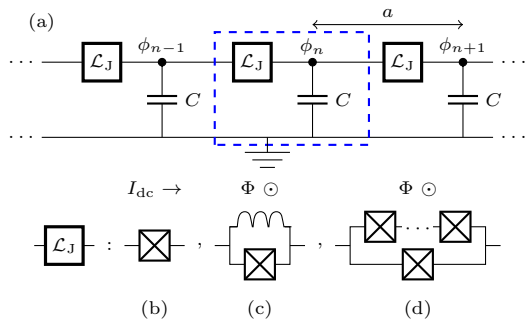
\begin{figure}[ht]
    \centering
    \begin{tikzpicture}\scriptsize

\draw (0,1) node[anchor=east]{\(\dots\)} -- (6,1) node[anchor=west]{\(\dots\)};
\draw (0,0) node[anchor=east]{\(\dots\)} -- (6,0) node[anchor=west]{\(\dots\)};

\foreach\x in {-1,0,1} {
    \draw[very thick,fill=white] (4+2*\x-1.45,0.75) rectangle(4+2*\x-0.95,1.25);
    \draw (4+2*\x-1.2,1) node{\(\mathcal L\J\)};
    \draw[fill] (4+2*\x-0.4,1) circle(0.05) node[anchor=south]{\(\phi_{n\signed\x}\)} -- +(0,-1);
    \capacitor{4+2*\x-0.4}{0.55}
}

\draw[dashed,blue,thick] (2.3,-0.1) rectangle(4.35,1.4);
\ground30
\draw (0,1.3) node[anchor=south]{(a)};
\draw[<->] (3.6,1.5) -- (5.6,1.5) node[midway,anchor=south]{\(a\)};

\draw (3,-1.5) node{\begin{tikzpicture}
    \draw (0.8,0) -- (1.6,0);
    \draw[very thick,fill=white] (0.95,-0.25) rectangle(1.45,0.25);
    \draw (1.2,0) node{\(\mathcal L\J\)};
    
    \draw(1.8,0) node{:};

    \draw (2,0) -- (2.8,0);
    \JJ{2.2}0
    \draw (2.4,0.75) node{\(I\dc\to\)};
    \draw (2.4,-0.85) node{(b)};
    
    \draw (3,0) node{,};
    
    \draw (3.2,0) -- (4.4,0);
    \draw[fill=white] (3.4,-0.25) rectangle(4.2,0.25);
    \hinductor[]{3.4}{0.25}
    \JJ{3.6}{-0.25}
    \draw (3.8,0.75) node{\(\Phi\ \odot\)};
    \draw (3.8,-0.85) node{(c)};
    
    \draw (4.6,0) node{,};
    
    \draw (4.8,0) -- (7,0);
    \draw[fill=white] (5,-0.25) rectangle(6.8,0.25);
    \fill[white] (5.7,0.2) rectangle(6.1,0.3);
    \draw (5.9,0.23) node{\(\cdots\)};
    \foreach\x in {5.2,6.2} {
        \JJ\x{0.25}
    }
    \JJ{5.7}{-0.25}
    \draw (5.9,0.75) node{\(\Phi\ \odot\)};
    \draw (5.9,-0.85) node{(d)};
    
\end{tikzpicture}};\end{tikzpicture}
    \caption{Circuit diagram of a general discrete TWPA (a), the dashed line indicates a unit cell of length \(a\). Options for the Josephson junction block \(\mathcal L\J\): (b) current biased junction, (c) flux-biased rf-SQUID, (d) flux-biased SNAIL.}
\label{fig:TWPAs}
\end{figure}

The travelling-wave parametric amplifier we study is a chain of identical cells, each consisting of a block of Josephson elements, \(\mathcal L\J\), and a capacitor, \(C\), as depicted in \Cref{fig:TWPAs}.

A convenient starting point for a dynamical description of the TWPA is the Lagrangian \cite{Devoret2004},
\begin{equation}
    \mathcal L = \left(\frac{\Phi_0}{2\pi}\right)^2 \,\sum_{n=1}^N \left( \frac C2 \,\dot\phi_n^2 + \mathcal L\J{} [\theta_n] \right) ,
\label{eq:general_L}
\end{equation}
where \(\phi_n(t)\) is the dynamical variable - the superconducting phase at node \(n\), \(\dot\phi_n(t)\) is its time derivative, \(\Phi_0 = h/(2e)\) is the magnetic flux quantum, \(\mathcal L\J\) is the Lagrangian of the Josephson junction block, and \(\theta_n = \phi_n -\phi_{n-1}\) is the superconducting phase difference across the block. 

We consider three flavours for the Josephson junction blocks suitable for 3WM.
The simplest one is the Josephson junction block consisting of a single Josephson junction, \Cref{fig:TWPAs}b,
\begin{equation}
    \mathcal L\J{} [\theta_n] = \left( \frac{C\J}2 \dot\theta_n^2 + \frac1{L\J}\cos\theta_{n} \right) ,
\label{eq:JJ_L}
\end{equation}
where \(C\J\) and \(L\J = \hbar/(2e I\crit)\) are the Josephson junction capacitance and inductance, respectively. 

In order to provide the 3WM mechanism, the Josephson junction has to be biased with a dc current, \(I\dc\), which induces a constant shift of the phase difference across each cell, \(\theta_0\), defined by equation,
\begin{equation}
    \sin\theta_0 = \frac{I\dc}{I\crit} \,.
\end{equation}

After including this constant shift to the phase difference, \(\theta_n \to \theta_0 +\theta_n(t)\), a dynamical equation for the TWPA is derived by varying the Lagrangian over the dynamical variable 
\(\phi_n(t)\) yielding,
\begin{equation}
\begin{aligned}
    &\ddot\phi_n - \frac{C\J}C (\ddot\theta_{n+1} - \ddot\theta_n) - \bar\omega_0^2 (\sin\theta_{n+1} - \sin\theta_n) \\
    &- \bar\omega_0^2\tan\theta_0 (\cos\theta_{n+1} - \cos\theta_n) = 0\,,
\end{aligned}
\label{eq:EOM_dc}
\end{equation}
where 
\begin{equation}
    \bar\omega_0^2 = \frac{\cos\theta_0}{L\J C}
\label{eq:omega0}
\end{equation}
is the resonance frequency of the current-biased \(L\J C\)-oscillator. 

The linear terms, with respect to \(\theta_n\), in \Cref{eq:EOM_dc} define the spectral properties of propagating waves in the TWPA, while the nonlinear terms, with respect to \(\theta_n\), are responsible for the mixing processes. In the absence of the biasing dc current the 3WM term vanishes. 

The TWPA dispersion relation is derived by assuming the solution to a linearised version of \Cref{eq:EOM_dc} as a discrete analogue to the propagating wave with quasi-wave vector \(\kappa\), \(\phi_n(t) \propto \e^{\I(\kappa n - \omega t)}\), giving the relation,
\begin{equation}
    2\bar\omega_0 \sin\frac\kappa2 = \frac\omega{\sqrt{1- (C\J/C)(\omega/\bar\omega_0)^2}} \,.
\label{eq:dispersion_dc}
\end{equation}
The dispersion relation has a cutoff at \(\kappa\crit = \pi\), where the frequency reaches the maximum value \(\omega\crit\),
\begin{equation}
    \omega\crit = \frac{2\bar\omega_0}{\sqrt{1+4C\J/C}}\,.
\label{eq:omega_cut}
\end{equation}
The dispersion relation cutoff is related to the discrete nature of the TWPA, but it is also affected by the Josephson junction capacitance. In practice, however, the latter is usually small, \(C\J \ll C\), and we will neglect it for most of the theoretical analysis throughout the paper, but keep it when comparing with experiments in \Cref{sec:CompWithExperiment}. 

Assuming a small amplitude of the phase oscillation, \(|\theta_n| \ll 1\), we expand trigonometric functions in \Cref{eq:EOM_dc} up to the second order, thus retaining the 3WM term but omitting the higher order 4WM term, to get,
\begin{equation}
    \ddot\phi_n - \bar\omega_0^2 (\theta_{n+1} - \theta_n) = -\frac{\bar\omega_0^2\tan\theta_0}2 \left( \theta_{n+1}^2 - \theta_n^2 \right) \,.
\label{eq:EOM_dc_simplified}
\end{equation}

An alternative solution for a 3WM TWPA is to use an rf-SQUID instead of a Josephson junction \cite{Zorin2016}, as shown in \Cref{fig:TWPAs}c. An rf-SQUID includes an inductance \(L\) in parallel with the Josephson junction, which allows replacing the dc current bias with a dc magnetic flux bias to achieve 3WM. The Lagrangian \(\mathcal L\J{}\) in this case has the form
\begin{equation}
    \mathcal L\J{} [\theta_n] = \left( \frac1{L\J} \cos(\theta_0 + \theta_n) - \frac{(\theta_0 +\theta_n + F)^2}{2L} \right) \,,
\end{equation}
where \(F = 2\pi\Phi/\Phi_0\) is the normalised magnetic flux.
The dc phase shift \(\theta_0\) is now defined, in the absence of net dc current through the rf-SQUID, by the relation
\begin{equation}
   \frac{\sin\theta_0}{L\J} + \frac{\theta_0 + F}L = 0 \,.
\end{equation}
The dynamic equation now takes a form similar to the one in \Cref{eq:EOM_dc_simplified},
\begin{equation}
\begin{aligned}
    & \ddot\phi_n - \omega^2\rf (\theta_{n+1} - \theta_n) = - \frac{\bar\omega_0^2\tan\theta_0}2 (\theta_{n+1}^2 - \theta_n^2)\,, \\
    & \omega^2\rf = \left(\bar\omega_0^2 + \frac1{LC}\right) \,.
\end{aligned}
\label{eq:EOM_rf}
\end{equation}

Another alternative design is to replace the rf-SQUID with a SNAIL circuit \cite{Sivak2019}, as shown in \Cref{fig:TWPAs}d. This device can be viewed as a modification of a dc-SQUID with several series connected junctions placed in one of the arms. The corresponding Lagrangian reads,
%
\begin{equation}
    \mathcal L\J{} [\theta_n] = \frac1{L\J[1]}\cos(\theta_0 +\theta_{n}+ F) - \frac{\mathcal N}{L\J[2]} \cos{\frac{\theta_0 +\theta_n}{\mathcal N}}, 
\label{eq:LSn}
\end{equation}
here \(\mathcal N\) refers to the number of identical junctions in the top arm in \Cref{fig:TWPAs}d. The biasing phase difference is defined in the absence of the net dc current through the SNAIL, by
\begin{equation}
    \frac{2\pi}{\Phi_0} \,I_S = \frac1{L\J[1]}\sin(\theta_0 + F) + \frac1{L\J[2]} \sin\frac{\theta_0}{\mathcal N} = 0 \,.
\label{eq:Delta_Sn}
\end{equation}
Proceeding with the derivation in a similar way to the single junction TWPA, we get the following dynamical equation for the SNAIL-TWPA,
\begin{equation}
    \ddot\phi_n - \omega_S^2 (\theta_{n+1} - \theta_n) = -\frac{\omega_S^2 \chi_3}2 \left( \theta_{n+1}^2 - \theta^2_n \right) \,, 
\label{eq:EOM_Sn}
\end{equation}
where
\begin{equation}
\begin{aligned}
    \omega^2_S &= \frac1{L\J[1]C}\cos(\theta_0 + F) + \frac1{L\J[2]C\mathcal N}\cos\frac{\theta_0}{\mathcal N} \,, \\
    \chi_3 &= \frac1{\omega^2_S} \left( \frac1{L\J[1]C}\sin(\theta_0 + F) + \frac1{L\J[2]C\mathcal N^2}\sin\frac{\theta_0}{\mathcal N} \right) .
\label{eq:omega_Sn}
\end{aligned}
\end{equation}

It is worth to note at this point that it is not possible to employ an asymmetric dc-SQUID for 3-wave mixing instead of the SNAIL: in this case \(\mathcal N=1\) in \Cref{eq:omega_Sn}, and the nonlinear term in \Cref{eq:EOM_Sn} turns to zero by virtue of \Cref{eq:Delta_Sn}.

Comparing \Cref{eq:EOM_dc,eq:EOM_rf,eq:EOM_Sn}, we find that they all have similar structures and could be written in a universal form,
\begin{equation}
\begin{aligned}
    &\frac1{\omega_0^2} \ddot\phi_n - ( \phi_{n+1} - 2\phi_n + \phi_{n-1} ) \\
    &= -\frac{\chi_3}2\left[ (\phi_{n+1} - \phi_n)^2 - (\phi_n - \phi_{n-1})^2 \right] \,, 
\end{aligned}
\label{eq:EOM_univers}
\end{equation}
where \(\omega_0\) is the resonance frequency of the cell, and \(\chi_3\) is the 3WM nonlinear coefficient. Particular values of these quantities for different TWPA designs are presented in \Cref{tab:c3_differentTWPAs}. 
\begin{table}[H]
    \centering
    \caption{Summary of resonance frequencies and 3WM nonlinear coefficients for different TWPA designs.}
    \label{tab:c3_differentTWPAs}
    \begin{tabular}{|c|c|c|c|}
        \hline
        TWPA & Bias & \(\omega_0\) & \(\chi_3\) \\ \hline
        Junction & Current & \(\bar\omega_0\) \Cref{eq:omega0} &\(\tan\theta_0\) \\ \hline
        RF-SQUID & Flux & \(\omega\rf\) \Cref{eq:EOM_rf} & \(\left(\frac{\bar\omega_0}{\omega\rf}\right)^2 \tan\theta_0\) \\ \hline
        SNAIL & Flux & \(\omega_S\) \Cref{eq:omega_Sn} & \Cref{eq:omega_Sn} \\ \hline
    \end{tabular}
\end{table}
%
\section{3-mode model}
\label{sec:Disc3modes}

The amplification in a TWPA results from the process of resonant down-conversion, when the frequencies of three interacting waves, \ie\ pump, signal, and idler, obey the resonance condition, \(\omega\pump = \omega\signal + \omega\idler\). Besides this process necessary for amplification, the nonlinear term in \Cref{eq:EOM_univers} generates a large set of up-converted modes of all the waves involved in the amplification and their combinations. These processes of up-conversion significantly degrade the performance of the amplifier, as we will show in the next section. 

To reveal the full amplification potential of the 3WM mechanism, we consider an ideal model where only three waves participating in the down-conversion are taken into account, while all the up-converted modes are neglected. In this model, the field in the TWPA chain consists of a linear combination of three partial harmonic tones,
\begin{equation}
    \phi_n(t) = \frac12 \sum_{\alpha=\text{p,s,i}} \left( A_{\alpha,n} \e^{-\I\omega_\alpha t} + \text{c.c.} \right)
\end{equation}
whose amplitudes satisfy equation that follows from  \Cref{eq:EOM_univers}, 
%
\begin{equation}
\begin{aligned}
    &\frac{\omega_\alpha^2}{\omega_0^2} A_{\alpha,n} + \left( A_{\alpha,n+1} - 2A_{\alpha,n} + A_{\alpha,n-1} \right) = \\
    &\frac{\chi_3\omega_\alpha}{2\pi} \int\displaylimits_0^{\frac{2\pi}{\omega_\alpha}} \left[ (\phi_{n+1}- \phi_n)^2 - (\phi_n - \phi_{n-1})^2 \right] \e^{\I\omega_\alpha t} \,\mathrm{d}t \,, 
\label{eq:A_alpha_eq}
\end{aligned}
\end{equation}
where integration is done over the period of the corresponding mode.
 
For the pump amplitude, the right-hand side of \Cref{eq:A_alpha_eq} only contains the resonant products, \(A\sAny{,n} A\iAny{,n}\), which are responsible for the pump depletion. For the amplifiers employed for qubit measurements, an input signal is typically \(\sim 1\)\,nA, which is by two orders of magnitude smaller than the pump current, which is typically a considerable fraction of the critical current, \(I\crit \sim 1\)\,\textmu A. Thus, for a power gain up to the order of 40 dB the signal and idler remain weak compared to the pump within the whole TWPA chain, \(A\sAny{,n}, \,A\iAny{,n} \ll A\pAny{,n}\), and we will neglect their effect on the pump. As a result, \Cref{eq:A_alpha_eq} for the pump becomes linear and has a free propagating wave solution, \(A\pAny{,n} = A\pump\e^{-\I\kappa\pump n}\), with the dispersion relation,
\begin{equation}
    \omega\pump^2 = 4\omega_0^2 \sin^2\frac{\kappa\pump}2\,.
\label{eq:spectrum} 
\end{equation}
The dispersion relation has a cutoff at \(\omega\crit= 2\omega_0,\; \kappa\crit = \pi\), and is strongly dispersive in the vicinity of the cutoff. In the long-wave limit, \(\kappa\pump\ll 1\), the dispersion relation becomes linear, \(\omega\pump \approx \omega_0 \kappa\pump \). 

Proceeding to the equations for signal and idler we find that the only resonant contributions here come from the products \(A\pAny{,n} A\iAny{,n}^*\) for the signal, and \(A\pAny{,n} A\sAny{,n}^*\) for the idler. This implies that the equations form a linear equation set. 
The solution ansatz has the form,
\begin{equation}
    A\sAny{,n} = A\signal \e^{\I\frac{\kappa\pump+\tilde\kappa}2 n}, \quad 
    A\iAny{,n}^* = A\idler^* \e^{-\I\frac{\kappa\pump-\tilde\kappa}2 n} \,,
\end{equation}
where \(\tilde\kappa\) is an unknown quasi-wave vector. Then the corresponding equations in \Cref{eq:A_alpha_eq} reduce to an algebraic equation for spatio-temporally independent amplitude, \(A\signal\), 
\begin{equation}
\begin{aligned}
    &\frac{\omega\signal^2}{\omega_0^2} A\signal - 4\sin^2 \frac{\kappa\pump + \tilde\kappa}4 A\signal = \\
    &= -\frac{\chi_3A\pump}{2} \left[ \left(1 - \e^{-\I\kappa\pump} \right) \left(1 - \e^{\I\frac{\kappa\pump - \tilde\kappa}2} \right) \right. \\
    &- \left. \left(\e^{\I\kappa\pump} - 1\right) \left( \e^{-\I\frac{\kappa\pump- \tilde\kappa}2} - 1 \right) \right] \,A\idler^* \,,
\end{aligned}
\label{eq:As_eqn}
\end{equation}
and a similar equation for \(A\idler\) with the replacements, \(A\signal \leftrightarrow A\idler\), \(\omega\signal \to \omega\idler\), and \(\kappa\pump \to -\kappa\pump\).

The quasi-wave vector \(\tilde\kappa\) is found from the solubility condition for \Cref{eq:As_eqn} and equation for \(A\idler\), \ie\ from the condition for the determinant of the system to be equal to zero. After some algebra the corresponding equation can be presented in the form,
\begin{equation}
\begin{aligned}
    0 &= \sin\frac{\kappa\pump - 2\kappa\signal + \tilde\kappa}4 \sin\frac{\kappa\pump - 2\kappa\idler - \tilde\kappa}4 \\
    &\times \sin\frac{\kappa\pump + 2\kappa\signal + \tilde\kappa}4 \sin\frac{\kappa\pump + 2\kappa\idler - \tilde\kappa}4 \\
    &- \chi_3^2|A\pump|^2 \sin^2\frac{\kappa\pump}2 \sin^2\frac{\kappa\pump + \tilde\kappa}4 \sin^2\frac{\kappa\pump - \tilde\kappa}4 \,.
\end{aligned}
\label{eq:Det_eqn}
\end{equation}
A solution to this equation generally has the imaginary part that describes the amplification effect. We quantify the amplification with the gain coefficient \(g = \Im (\tilde\kappa/2)\), which characterises the amplitude gain per unit cell and is related to the power gain of TWPA as
\begin{equation}
    G(\text{dB}) = 20gN\log_{10}(\e)\,,
\end{equation}
where \(N\) is the number of unit cells. 
 
\begin{figure}[H]
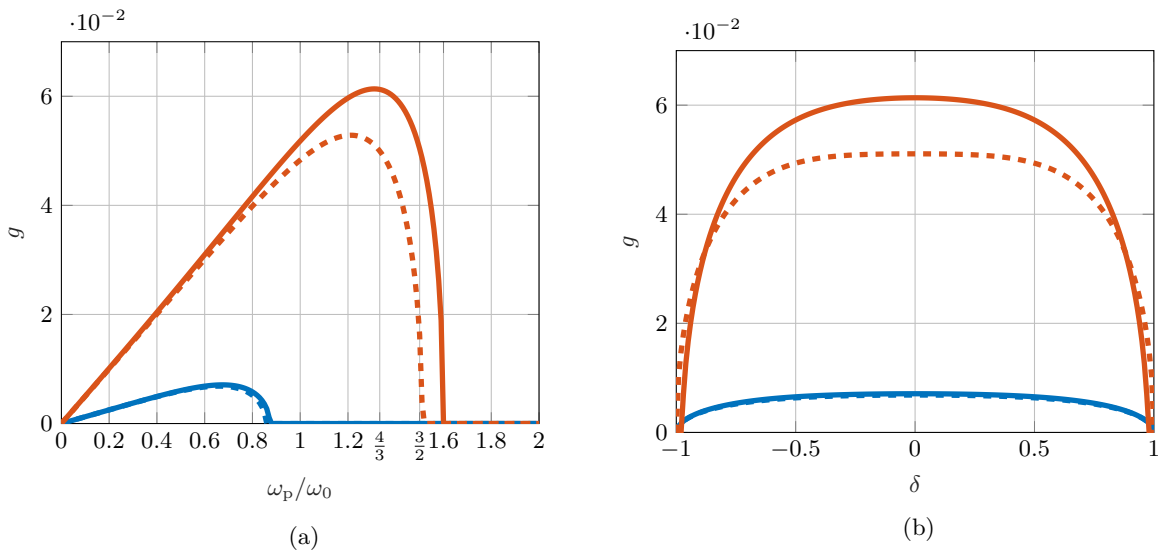

    \centering
    \begin{subfigure}{0.45\textwidth}
        \begin{tikzpicture}\input{tikzpics/IncrementVsPumpFreq/IncrementVsFp_VitDfVsApproxVsLin_delta0_chi_0p025_and_0p1.tikz}\end{tikzpicture}
        \caption{}
        \label{fig:IncrementVsPumpFreq_Ap}
    \end{subfigure}
    \begin{subfigure}{0.45\textwidth}
        \begin{tikzpicture}\input{tikzpics/IncrementVsDelta/IncrementVsDelta_VitDfVsApproxVsLin_fpmax_chi_0p025_and_0p1.tikz}\end{tikzpicture}
        \caption{}
        \label{fig:g_vs_delta}
    \end{subfigure}
    \caption{Gain coefficient from numerically solving \Cref{eq:Det_eqn} (solid), and from \Cref{eq:g_solution} (dashed), for different pumping strengths, \(\varepsilon = 0.1\) (blue) and 0.4 (orange). (a) Gain coefficient at zero detuning, \(\delta=0\), as function of pump frequency. (b) Gain coefficient as function of detuning at the pump frequencies, \(\omega\pump/\omega_0 \) = 0.67 and 1.31, corresponding to the maximum gain coefficient at zero detuning.}
    \label{fig:g_vs_pumpfreq_and_delta}
\end{figure}

\begin{figure}[ht]
    \centering
    \includegraphics[width=\linewidth]{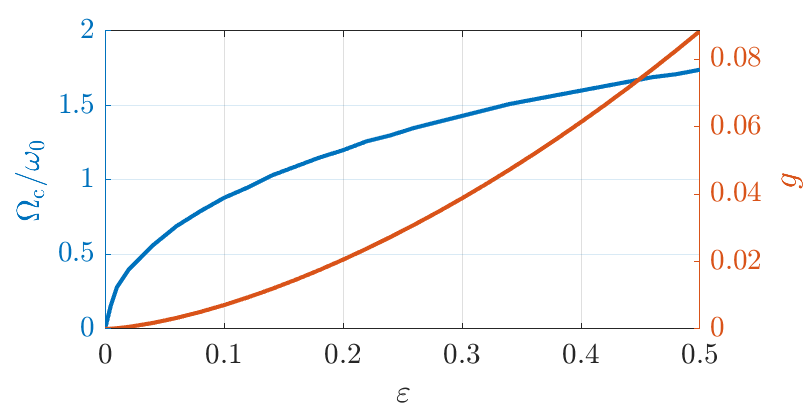}
    \caption{Maximum value of the gain coefficient (orange) and the gain coefficient cutoff frequency \(\Omega\crit\) (blue)  as functions of the pumping strength.}
    \label{fig:gcut_vs_gmax}
\end{figure}

The exact numerical solution for \(g\) at the degeneracy point, \(\omega\signal = \omega\idler = \omega\pump/2\), is presented in \Cref{fig:IncrementVsPumpFreq_Ap} for different pumping strengths. The pumping strength here is characterised through the quantity
\begin{equation}
    \varepsilon = |\chi_3\theta\pump|\,, \quad
    \theta\pump = 2 \sin\frac{\kappa\pump}2 A\pump = \frac{\omega\pump}{\omega_0}A\pump\,.
\label{eq:epsilonAp}
\end{equation}
Here \(\theta\pump\) is the the amplitude of oscillation of phase difference across the cell associated with the 
pump amplitude at the node, \(A\pump\). At small frequencies, \(\omega\pump \ll \omega\crit\), \(\kappa\pump\ll1\), the gain coefficient is small and grows linearly with frequency. It reaches a maximum and then sharply drops to zero at the gain cutoff frequency, \(\omega\pump = \Omega\crit\). The maximum gain and \(\Omega\crit\) 
depend on the pumping strength as illustrated in \Cref{fig:gcut_vs_gmax}. As seen from \Cref{fig:IncrementVsPumpFreq_Ap}, the maximum gain at large frequencies is quite large: for pumping strength \(\varepsilon \approx 0.4\), the gain coefficient reaches the value \(g \approx 0.06\), which translates to the power gain, \(G \approx 26\,\text{dB}\), for a chain with \(N=50\) cells.

A numerical solution of \Cref{eq:Det_eqn} for the gain coefficient as a function of detuning is presented in \Cref{fig:g_vs_delta} with solid lines. The plots are made for the pump frequency values corresponding to maxima of the gain coefficients in \Cref{fig:IncrementVsPumpFreq_Ap} for the same pump intensities. 
 
To better understand the gain properties, we derive an approximate analytical solution to \Cref{eq:Det_eqn}. To this end we cast \(\tilde\kappa\) on the form, 
\begin{equation}
    \tilde\kappa = \kappa\signal - \kappa\idler - 2\I g \,,
\end{equation}
where quasi-wave vectors, \(\kappa_\text{s,i}\), are related to the signal and idler respective frequencies via the dispersion relation similar to \Cref{eq:spectrum}. We also define the detuning from the degeneracy point,
\begin{equation}
    \delta = \frac{\omega\signal - \omega\pump/2}{\omega\pump/2} \,,
\label{eq:delta}
\end{equation}
and the phase mismatch \(\Delta(\delta)\),
\begin{equation}
    \Delta(\delta) = \kappa\pump - \kappa\signal(\delta) - \kappa\idler(\delta) \, .
\label{eq:Delta}
\end{equation}
At weak coupling, \(\varepsilon \ll1\), the gain coefficient is small, as is also seen in \Cref{fig:IncrementVsPumpFreq_Ap}, in comparison with the wave vectors \(\kappa\pump \sim \kappa_\mathrm{s} \sim \kappa_\mathrm{i}\), which are of order unity at high frequencies, as given by the dispersion relation in \Cref{eq:dispersion_dc}. On the other hand, the phase mismatch, which is sufficient to suppress the gain is also small. Therefore, both \(g\) and \(\Delta(\delta)\) are small additive corrections to \(\kappa_\mathrm{p,s,i}\), and can be omitted from the corresponding terms in \Cref{eq:Det_eqn}. This approximation yields the explicit solution for \(g\),
\begin{equation}
\begin{aligned}
    \sinh^2\frac g2 &= -\sin^2\frac{\Delta(\delta)}4 + \chi_3^2|A\pump|^2 \frac{\sin^2\frac{\kappa\pump}2}{\sin \kappa\signal \sin \kappa\idler} \\
    &\times \left(\sin^2\frac{\kappa\signal + \kappa\idler}4 - \sin^2\frac{\kappa\signal-\kappa\idler}4 \right)^2 \,. \\
    &\quad
\end{aligned}
\label{eq:g_solution}
\end{equation}
The obtained solution is a generalisation to a discrete chain and arbitrary frequency of the result obtained in \cite{Tien1958} for a continuous medium.
One can see from \Cref{eq:g_solution} that it is the competition between the nonlinear coupling controlled by the intensity of the pump (the second term on the right) and the phase mismatch (the first term on the right) that determines whether \(g\) is real or imaginary, {\it i.e.}, whether exponential amplification occurs or not. The sharp drop of the gain is explained by the increasingly strong dispersion near the cutoff frequency.
At zero detuning, \Cref{eq:g_solution} reduces to the form
\begin{equation}
    \sinh^2\frac g2 = - \sin^2\frac\Delta4 + \chi_3^2|A\pump|^2 \sin^4\frac{\kappa\pump}4 \,,
\label{eq:g_degen}
\end{equation}
where 
\begin{equation}
    \Delta = \Delta(0) = \kappa\pump - 2\kappa\signal
\label{eq:Delta0}
\end{equation}
is the phase mismatch between the frequency points, \(\omega\pump\) and \(\omega\pump/2\). This solution is represented in \Cref{fig:g_vs_pumpfreq_and_delta} with dashed lines.

Equation (\ref{eq:Det_eqn}) and its approximate analytical solution, \Cref{eq:g_solution}, together with numerical solution presented in \Cref{fig:g_vs_pumpfreq_and_delta}
constitute the first main result of this paper.

\onecolumngrid

\vspace{1cm}

\twocolumngrid

%

\subsection{Validity of the model}
\label{sec:validity}
\begin{figure}[ht]
    \centering
    \begin{tikzpicture}
\def\xtickdata{0 3 6}
\readarray\xtickdata\xticks[1,3]
\def\xticklabeldata{0 \(\frac12\) 1}
\readarray\xticklabeldata\xticklabels[1,3]
\def\ytickdata{0 3 4 6}
\readarray\ytickdata\yticks[1,4]
\def\yticklabeldata{0 1 \(\frac43\) 2}
\readarray\yticklabeldata\yticklabels[1,4]

\draw[->] (0,0) -- (6.5,0) node[anchor=west]{\(\omega\signal/\omega\pump\)};
\draw[->] (0,0) -- (0,6.5) node[anchor=south]{\(\omega\pump/\omega_0\)};
\foreach\x in {1,2,3} { 
    \draw (\xticks[1,\x],0) -- +(0,-0.1) node[anchor=north]{\xticklabels[1,\x]};
}
\foreach\y in {1,2,3,4} { 
    \draw (0,\yticks[1,\y]) -- +(-0.1,0) node[anchor=east]{\yticklabels[1,\y]};
}

\fill[red,opacity=0.25] (0,0) rectangle(6,3);
\draw (3,1.5) node{p, s, i};

\fill[yellow, opacity=0.25, domain=0:1, smooth, variable=\x] plot(6*\x,{6/(1+\x)}) -- (0,3) -- (0,6);
\draw (1,4.5) node{s};

\fill[yellow, opacity=0.25, domain=0:1, smooth, variable=\x] plot(6*\x,{6/(2-\x)}) -- (6,3) -- (0,3);
\draw (5,4.5) node{i};

\draw[dashed, thick, domain=0:1, smooth, variable=\x] plot(6*\x,{6/(1+\x)});
\draw[dashed, thick, domain=0:1, smooth, variable=\x] plot(6*\x,{6/(2-\x)});
\draw[dashed, thick] plot(0,4) -- (6,4);

\draw (3,3.35) node{s, i};

\fill[green,opacity=0.25,domain=0:0.5,smooth,variable=\x] plot(6*\x,{6/(1+\x)}) -- plot({6*(\x+0.5)},{6/(2-\x-0.5)}) -- (0,6);
\draw (3,5.25) node{No up-conversion};

\end{tikzpicture}
    \caption{The regions of the absence/presence of up-conversion of pump, signal and idler. In the green region no up-conversion takes place, horizontal dashed line indicates \(\Omega_\text{th}\); in light yellow regions up-conversion of either signal or idler is possible, while in yellow region both signal and idler are up-converted but pump is not; in pink region all three modes are up-converted.}
    \label{fig:UpconversionDisabling}
\end{figure}
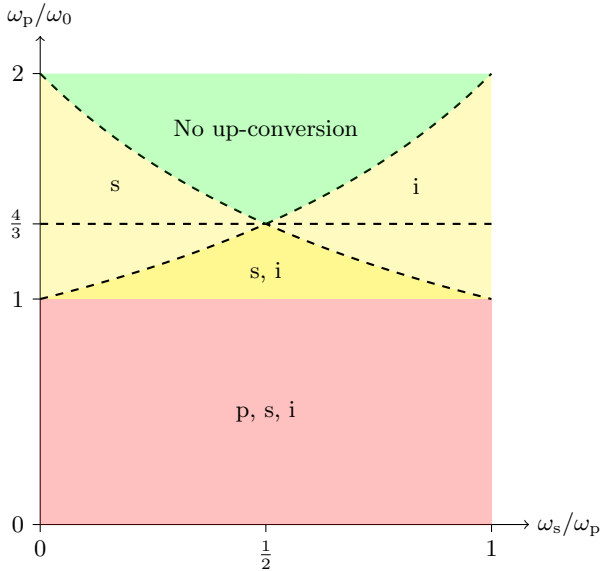

The validity of the model above relies on the absence of up-converted modes of the pump, signal, and idler. For the pump, the condition, \(\omega\pump > \omega_0\), guarantees that the second pump harmonic falls above the cutoff, \(\omega\crit = 2\omega_0\). For the signal/idler the lowest bound is established by condition that the up-converted signal at zero detuning falls above the cutoff, \(\omega\pump/2 + \omega\pump > 2\omega_0\). This yields a more stringent constraint, \(\omega\pump > \Omega_\text{th} = 4\omega_0/3\). At pump frequency larger than this threshold value, the detuned signal and idler are not up-converted within the band defined by equation,
\begin{equation}
    |\delta| < 3\left( 1 - \frac{\Omega_\text{th}}{\omega\pump} \right).
\label{eq:UpconversionDisabledBand}
\end{equation}
This situation is illustrated in \Cref{fig:UpconversionDisabling}. 

Therefore we conclude that the three-mode model considered is justified, when the gain cutoff frequency exceeds the threshold for no-up-conversion, \(\Omega\crit(\varepsilon) > \Omega_\text{th}\), and within the bandwidth in \Cref{eq:UpconversionDisabledBand}. This condition imposes the lowest bound for the required pumping strength. An accurate estimate of the lowest bound is extracted from the numerical solution to \Cref{eq:Det_eqn}, 
\begin{equation}
    \varepsilon \gtrsim 0.25 \,.
\label{eq:maxepsilon}
\end{equation}
To obtain an analytical estimate we assume, \(g=0\) in \Cref{eq:g_degen}, to get,
\begin{equation}
    \chi_3|A\pump| > \frac{\sin\frac\Delta4}{\sin^2\frac{\kappa\pump}4} \,.
\label{eq:Ap_max}
\end{equation}
At \(\omega\pump = \Omega_\text{th}\), \(\kappa\pump \approx 0.46\pi\), and for zero detuning, \(\delta=0\), we have \(\kappa\signal \approx 0.22\pi\), giving \(\Delta \approx 0.03\pi\). This results in the bound for the pumping strength, \(\chi_3|A\pump| > 0.20\), or \(\varepsilon \gtrsim 0.27\), which slightly overestimates the exact result from \Cref{eq:maxepsilon}. 

The crucial question is now whether the required coupling strength can be  experimentally achieved with a feasible pump intensity. 
Let us first consider the dc current biased TWPA. The pumping strength here is limited by the switching of the Josephson junctions to the resistive branch. In the quantum limit, the maximum dc supercurrent that the junction can sustain corresponds to the disappearance of the last quantized energy level from the well of the tilted Josephson potential. 
The maximum dc supercurrent can be crudely estimated from the relation, \(\hbar\omega_\text{pl} /2 \sim \Delta U\), where \(\omega_\text{pl} = \sqrt{\cos\theta_0/(L\J C_\Sigma)}\) is the effective plasma frequency for the junction, \(C_\Sigma = C/2 + C\J \approx C/2\), and \(\Delta U \approx (2E\J/3)\cos^3\theta_0\) is the depth of the well of the tilted Josephson potential. Impedance matching of the TWPA with the transmission line, \(\sqrt{L\J/(C\cos\theta_0)} = Z_0 \), gives for the maximum current, \(\cos^2\theta_0 \sim 3\pi \sqrt2 Z_0/R_q\), where \(R_q = h/(2e^2)\) is the quantum resistance; this corresponds to \(I\dc \sim 0.97I\crit\). Assuming that the quasi-classical tunnelling rate, \(\Gamma\propto \exp( -7.2\Delta U/\hbar\omega_\text{pl})\) \cite{Caldeira1983}, is valid for the two-three quantized energy levels in the well \cite{Hanggi1987}, and also taking into account the experimental observations, \eg\ in Ref.~\cite{Yu2010}, we may safely assume for the switching current value, \(I\approx 0.9I\crit\).

When the pump is on and has a small frequency, \(\omega\pump \ll \omega_\text{pl}\), the instant adiabatic current consists of the dc biasing current, \(I\dc\), and the pump ac current, \(I\pump\), and their sum should not exceed the switching current, \(I\dc + I\pump < 0.9I\crit\). The maximum pumping strength under this constraint is, \(\varepsilon = 0.28\), which is achieved at \(I\dc = 0.63I\crit\) and \(I\pump = 0.27I\crit\). Although this pumping strength is above the bound in \Cref{eq:maxepsilon}, the frequency window where the model is valid is very small, \(\Omega\crit - \Omega_\text{th} \approx 0.06\omega_0\), \Cref{fig:UpconversionDisabling}, making the amplification bandwidth unacceptably narrow. Furthermore, the corresponding pumping current is too large given the theory constraint, \(I\pump\ll I\crit\), more feasible would be the lower current values, \(I\pump \sim 0.1I\crit\). In addition, a spread of the junction parameters in a real TWPA would also reduce the estimated pumping strength.

More importantly, however, is that the relevant ac regime is non-adiabatic: the pump frequencies above the no-up-conversion threshold, \(\omega\pump > \Omega_\text{th} \approx 1.33\bar\omega_0\), are close to and even higher than the plasma frequency, \(\omega_\text{pl} = \sqrt2 \bar\omega_0 \approx 1.41\bar\omega_0\). In this regime, the resonant excitation facilitates tunnelling (especially due to the multi-photon processes at large pump amplitude), therefore the biasing current should be even smaller than in the adiabatic regime, hence the pumping strength would be further reduced. 

In the case of an rf-SQUID TWPA, the maximum nonlinear coupling is achieved at \(\theta_0 = \pi/2\) \cite{Zorin2016}, when \(\chi_3 = L/L\J\). For a non-hysteretic regime, \(L/L\J<1\), combination of this constraint with the a small value of the amplitude of phase oscillation, \(\theta\pump < 0.1\), results in a pumping strength, \(\varepsilon < 0.1\), which is below the threshold, \Cref{eq:maxepsilon}. Similar argument also applies to the SNAIL-TWPA.

To summarise, we conclude that the 3-mode amplification regime, which would provide the high exponential gain at high frequencies, \(\omega\pump \sim \omega_0\)  cannot be realised in practice with any of the TWPA designs considered here. The desired condition, \(\Omega\crit(\varepsilon) > \Omega_\text{th}\), cannot be fulfilled  because of small values of nonlinear 3WM coefficients in realistic devices and limited pumping current. 

To go beyond the studied 3-mode model, two strategies can be followed. The one is to consider the lower frequencies and include the up-converted modes in the model, while the other is to keep the 3-mode model but consider dispersion engineering at high frequencies. In the next section we will discuss the first option in detail.

\section{Quasilinear dispersion regime}
\label{sec:quasilinear}
In this section we analyse TWPA performance in the low frequency region, \(\omega\ll\omega_0, \; \kappa\ll\pi\), where the dispersion relation is quasi-linear. The limit of the continuous medium is natural for the kinetic inductance TWPA, but it is also considered in most of publications devoted to the Josephson junction TWPAs.

In this limit the discrete chain of the Josephson junctions is described with a continuous variable, \(an \to x, \;\phi_n \to \phi(x)\), where \(a\) is the physical length of the unit cell. Then the difference equation \Cref{eq:EOM_univers} turns into a differential equation, 
\begin{equation}
\begin{aligned}
    &\frac1{\omega_0^2}\ddot\phi + 4\sin^2\frac{a\hat k}2\phi \\
    &= \frac{\chi_3}2 \left\{ \left[ \left(1 - \e^{-\I a\hat k} \right) \phi \right]^2 - \left[ \left( \e^{\I a\hat k} - 1 \right) \phi \right]^2 \right\}\,,
\end{aligned}
\label{eq:cont_eqn}
\end{equation}
where \(\hat k = -\I\partial_x\).
Keeping the lowest order terms with respect to \(a\) in the expansion of the interaction term, we write \Cref{eq:cont_eqn} in the form, 
%
%
%
\begin{equation}
    \frac1{\omega_0^2}\ddot\phi + 4\sin^2\frac{a\hat k}2 \phi \, = \,  \frac{\chi_3a^3}2 \,i\hat k ( \hat k\phi )^2  \,. 
\label{eq:diff_eqn}
\end{equation}
This dynamical equation (and a similar one for the 4WM) is a standard object of study in the TWPA literature. 

To analyse the resonant wave dynamics described by this equation one has to take into account, in addition to the down-conversion discussed in the previous section, the processes which are efficient for a weakly dispersive medium: (i) generation of pump harmonics with frequencies \(n\omega\pump\) \cite{Bloembergen1962}, (ii) up-conversion of the signal and down-converted idler by the pump tone and its harmonics, \(\omega_\text{s,i}+n\omega\pump\) \cite{Dixon2020}. The processes that can be neglected for a weak signal include pump depletion and generation of signal/idler harmonics and their intermodulation products.

\subsection{Pump harmonics}
\label{sec:3WMPumpHarmonics}

Let us first consider pump harmonic generation. When a pump tone is injected, the field in the cavity has the form
\begin{equation}
    \phi(x,t) = \frac12 \sum_{m=1}^M \left(A\ph m(x) \e^{\I(k\ph mx - m\omega\pump t)} + \text{c.c.}\right) \,,
\label{eq:pump}
\end{equation}
where \(k\) is a wave vector related to \(\omega\) via the free wave dispersion relation, \(ka = 2\arcsin(\omega/(2\omega_0))\), and \(M\) is the number of harmonics included in the computation (see below). A slow variation of the amplitudes of the harmonics accounts for the effect of nonlinear interaction. 

\begin{figure}[ht]
    \centering
    \begin{tikzpicture}
\renewcommand\freqscale{1.7}
\renewcommand\mixheightlower{0.75}
\renewcommand\mixheightupper{0.5}
\draw[->] (-4.25,0) -- (4.25,0);
\foreach\x in {-2,-1,0,1,2} {
    \setcounter{ph}\x
    \draw[very thick, -stealth] (\freqscale*\x,0) -- +(0,0.7);
    \draw (\freqscale*\x,0) node[anchor=north]{
        \ifnum \value{ph}=0 {\textcolor{white}{(}\(m\)p\textcolor{white}{)}} \fi
        \ifnum \value{ph}>0 {\((m+\theph)\)p} \fi
        \ifnum \value{ph}<0 {\((m\theph)\)p} \fi
    };
}
\foreach\h in {1,2} {
    \draw (0.05,\mixheightlower) to[out=45+5*\h,in=180] (0.5*\freqscale*\h,\mixheightlower+\h*\mixheightupper) node[anchor=north]{\ifnum\h>1{\h}\fi p} to[out=0,in=135-5*\h] (\freqscale*\h-0.05,\mixheightlower);
    \draw[blue] (-\freqscale*\h+0.05,\mixheightlower) to[out=45+5*\h,in=180] (-0.5*\freqscale*\h,\mixheightlower+\h*\mixheightupper) node[anchor=north]{\ifnum\h>1{\h}\fi p} to[out=0,in=135-5*\h] (-0.05,\mixheightlower);
}
\draw (0.05,\mixheightlower) to[out=45+5*3,in=180] (0.5*\freqscale*3,\mixheightlower+3*\mixheightupper) node[anchor=north]{3p} to[out=0,in=165] (\freqscale*2,\mixheightlower+2.78*\mixheightupper);
\draw[dotted] (\freqscale*2,\mixheightlower+2.78*\mixheightupper) to[out=165+180,in=150] (\freqscale*2+0.75,\mixheightlower+2.07*\mixheightupper);
\draw[blue] (-0.05,\mixheightlower) to[out=180-45-5*3,in=180-180] (-0.5*\freqscale*3,\mixheightlower+3*\mixheightupper) node[anchor=north]{3p} to[out=180-0,in=180-165] (-\freqscale*2,\mixheightlower+2.78*\mixheightupper);
\draw[dotted, blue] (-\freqscale*2,\mixheightlower+2.78*\mixheightupper) to[out=180-165-180,in=180-150] (-\freqscale*2-0.75,\mixheightlower+2.07*\mixheightupper);

\end{tikzpicture}
    \caption{Interaction of \(m\)-th pump harmonic illustrating structure of \Cref{eq:Eq_Ap}. 
  Each process is marked with the pump harmonic that is driving the process.}
\label{fig:PumpHarmonicsIllustration}
\end{figure}
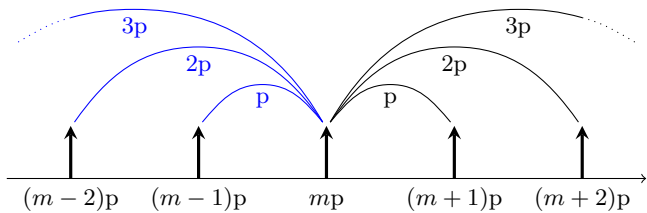

Substituting \Cref{eq:pump} in \Cref{eq:diff_eqn} we get a set of \(M\) coupled nonlinear equations for the pump harmonics,
\begin{equation}
\begin{aligned}
    &A'\ph m = \frac{\chi_3a}4 \times \\
    &\left(\sum_{n=m+1}^M k\ph n k\ph{(n-m)} A\ph n A\ph{(n-m)}^* \e^{\I\Delta_{n,m} x} \right. \\
    &\left.- \frac12 \sum_{n=1}^{m-1} k\ph n k\ph{(m-n)} A\ph n A\ph{(m-n)} \e^{-\I\Delta_{m,n} x} \right),
\label{eq:Eq_Ap}
\end{aligned}
\end{equation}
where the prime signifies spatial derivative, \(A' = \rd A/ \rd x\), \(\Delta_{n,m} = k\ph n - k\ph m - k\ph{(n-m)}\), and \(m\in\{1,...,M\}\) is the harmonic number. 
In this equation, the first sum describes coupling to the higher harmonics, while the second sum describes coupling to the lower harmonics, as illustrated in \Cref{fig:PumpHarmonicsIllustration}. The factor 1/2 in the second sum accounts for the double counting in the sum, \eg\, \(n=1\) and \(n=m-1\). 

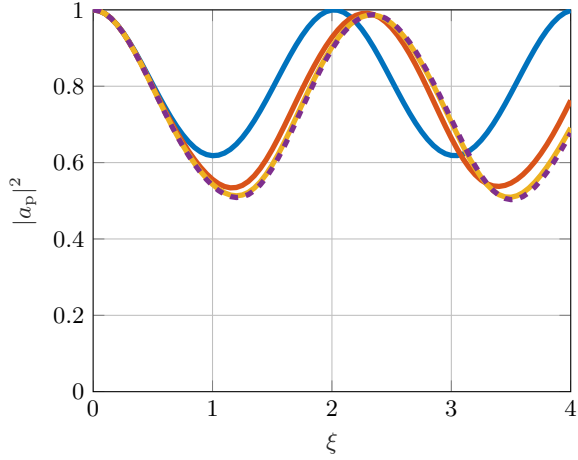
\begin{figure}[ht]
    \centering
    \begin{tikzpicture}
%
%
\definecolor{mycolor1}{rgb}{0.00000,0.44700,0.74100}%
\definecolor{mycolor2}{rgb}{0.85000,0.32500,0.09800}%
\definecolor{mycolor3}{rgb}{0.92900,0.69400,0.12500}%
\definecolor{mycolor4}{rgb}{0.49400,0.18400,0.55600}%
%

\begin{axis}[%
width=2.5in,
height=2in,
at={(2.6in,1.103in)},
scale only axis,
xmin=0,
xmax=4,
xlabel style={font=\color{white!15!black}},
xlabel={\(\xi\)},
ymin=0,
ymax=1,
ylabel style={font=\color{white!15!black}},
ylabel={\(|a\pump|^2\)},
axis background/.style={fill=white},
xmajorgrids,
ymajorgrids,
legend style={at={(0.03,0.03)}, anchor=south west, legend cell align=left, align=left, draw=white!15!black}
]
\addplot [color=mycolor1, line width=2.0pt, forget plot]
  table[row sep=crcr]{%
0	1\\
0.05	0.997506241902454\\
0.1	0.990099314358531\\
0.15	0.977999650669364\\
0.2	0.961562810794613\\
0.25	0.94126458366544\\
0.3	0.917683404164322\\
0.35	0.891475813717628\\
0.4	0.863348286615433\\
0.45	0.834040542508442\\
0.5	0.804298062878004\\
0.55	0.774849566579084\\
0.6	0.746390163182239\\
0.65	0.719569436721181\\
0.7	0.694977101902163\\
0.75	0.673133354797962\\
0.8	0.654486267778655\\
0.85	0.639407404992018\\
0.9	0.628189634277766\\
0.95	0.621045781897327\\
1	0.618109687625297\\
1.05	0.619435998246928\\
1.1	0.625000332661477\\
1.15	0.634698884089794\\
1.2	0.648348117815306\\
1.25	0.665684994655991\\
1.3	0.686368184920813\\
1.35	0.709979336497878\\
1.4	0.736023827347067\\
1.45	0.763936035073667\\
1.5	0.793089278713962\\
1.55	0.822805785562287\\
1.6	0.852365789808076\\
1.65	0.881025911409402\\
1.7	0.908044327720982\\
1.75	0.932699707089163\\
1.8	0.954312258690004\\
1.85	0.972270086594682\\
1.9	0.986058221130675\\
1.95	0.995273394066919\\
2	0.999641939158245\\
2.05	0.999032834211247\\
2.1	0.993465078311556\\
2.15	0.983104895326962\\
2.2	0.968258491326984\\
2.25	0.949357081057135\\
2.3	0.926943738932259\\
2.35	0.901648356818677\\
2.4	0.874161523096972\\
2.45	0.845212261966691\\
2.5	0.815546936606678\\
2.55	0.785903726600569\\
2.6	0.756991248043548\\
2.65	0.729478340436097\\
2.7	0.703978365586385\\
2.75	0.681038923740678\\
2.8	0.661135324800627\\
2.85	0.644668627287515\\
2.9	0.631961633933828\\
2.95	0.623257469101657\\
3	0.618719734996293\\
3.05	0.618433007051373\\
3.1	0.622402772299628\\
3.15	0.630555368597767\\
3.2	0.642737356956515\\
3.25	0.658715949853475\\
3.3	0.678178778289748\\
3.35	0.700736797102423\\
3.4	0.725923718762056\\
3.45	0.753200659985371\\
3.5	0.78196155542946\\
3.55	0.811547579399081\\
3.6	0.841252911252152\\
3.65	0.870341496526772\\
3.7	0.898066661861588\\
3.75	0.923698231226466\\
3.8	0.946537254187398\\
3.85	0.96594335503357\\
3.9	0.981361673197026\\
3.95	0.992345330615075\\
4	0.998569811301093\\
};

\addplot [color=mycolor2, line width=2.0pt, forget plot]
  table[row sep=crcr]{%
0	1\\
0.05	0.997504698092651\\
0.1	0.990075600189979\\
0.15	0.97788606057696\\
0.2	0.961232635616678\\
0.25	0.940539871993647\\
0.3	0.916356845074021\\
0.35	0.889339466722406\\
0.4	0.860217514578882\\
0.45	0.829750973387875\\
0.5	0.798678805874763\\
0.55	0.767673208430515\\
0.6	0.737306473475051\\
0.65	0.708033992084922\\
0.7	0.680199265719993\\
0.75	0.654057436463781\\
0.8	0.629809783545894\\
0.85	0.607644631470391\\
0.9	0.587775215667804\\
0.95	0.570464517787833\\
1	0.556035582018969\\
1.05	0.544863571312157\\
1.1	0.537350278560966\\
1.15	0.533886239563167\\
1.2	0.534806656687661\\
1.25	0.540346378231798\\
1.3	0.550600130030869\\
1.35	0.565494565479314\\
1.4	0.584775045169182\\
1.45	0.608008557739797\\
1.5	0.634604866288234\\
1.55	0.663854128314579\\
1.6	0.694974100889658\\
1.65	0.727163254535693\\
1.7	0.759651546549609\\
1.75	0.791738444144879\\
1.8	0.822815294484625\\
1.85	0.852369330153936\\
1.9	0.879970292731878\\
1.95	0.90524687366508\\
2	0.927862362276714\\
2.05	0.947496246123814\\
2.1	0.96383748388665\\
2.15	0.976592121425211\\
2.2	0.985502041755821\\
2.25	0.990366254613809\\
2.3	0.991061012662136\\
2.35	0.987550084030193\\
2.4	0.979882950022738\\
2.45	0.968183642140026\\
2.5	0.95263450007252\\
2.55	0.933462627250411\\
2.6	0.910935024284336\\
2.65	0.885363919523518\\
2.7	0.857123827395695\\
2.75	0.826672508695825\\
2.8	0.794567630675734\\
2.85	0.761473204969016\\
2.9	0.728148834878271\\
2.95	0.695420022149846\\
3	0.664133481541035\\
3.05	0.635104984788587\\
3.1	0.609065700491259\\
3.15	0.586616230092657\\
3.2	0.568196064838723\\
3.25	0.554069118705419\\
3.3	0.544327098420716\\
3.35	0.538910802657103\\
3.4	0.537643765355214\\
3.45	0.540272874879027\\
3.5	0.546513040442197\\
3.55	0.55608701163033\\
3.6	0.568754457660353\\
3.65	0.584327907681066\\
3.7	0.602671004251704\\
3.75	0.623679675377416\\
3.8	0.647249493613057\\
3.85	0.673236046187529\\
3.9	0.701413855177082\\
3.95	0.731442315126359\\
4	0.762847284325188\\
};

\addplot [color=mycolor3, line width=2.0pt, forget plot]
  table[row sep=crcr]{%
0	1\\
0.05	0.99750468835739\\
0.1	0.99007506502447\\
0.15	0.977880596386734\\
0.2	0.961205794438197\\
0.25	0.940452100035521\\
0.3	0.916134144248859\\
0.35	0.888863582690886\\
0.4	0.859320050669621\\
0.45	0.828216639740819\\
0.5	0.796269029944779\\
0.55	0.764170077615985\\
0.6	0.732562862291348\\
0.65	0.702008604041283\\
0.7	0.672956249818286\\
0.75	0.645729336914404\\
0.8	0.620542748040294\\
0.85	0.597543943077531\\
0.9	0.57686217128423\\
0.95	0.558646105889374\\
1	0.54308503396657\\
1.05	0.53041813387487\\
1.1	0.520939256118776\\
1.15	0.514992750163668\\
1.2	0.512950917192102\\
1.25	0.515166104348434\\
1.3	0.521907204217152\\
1.35	0.533300487443046\\
1.4	0.549294715934063\\
1.45	0.569655871709644\\
1.5	0.593981583818231\\
1.55	0.621722312029099\\
1.6	0.652202640859308\\
1.65	0.684650209675285\\
1.7	0.71823874994785\\
1.75	0.752144240933065\\
1.8	0.78559784269111\\
1.85	0.817918768973075\\
1.9	0.848521450290758\\
1.95	0.876906897896678\\
2	0.902652580288149\\
2.05	0.92540721499357\\
2.1	0.944885697601878\\
2.15	0.960859281836318\\
2.2	0.973142614771656\\
2.25	0.981586544571213\\
2.3	0.986081864614467\\
2.35	0.986571558040821\\
2.4	0.983063957817298\\
2.45	0.97564042389528\\
2.5	0.964454903092495\\
2.55	0.949724130827942\\
2.6	0.931708642953443\\
2.65	0.910689348947741\\
2.7	0.886950119382558\\
2.75	0.86077702698818\\
2.8	0.832475317116187\\
2.85	0.80239331884093\\
2.9	0.770940902787221\\
2.95	0.738597396654164\\
3	0.705915445607795\\
3.05	0.673525413363832\\
3.1	0.642135993131605\\
3.15	0.612516682506347\\
3.2	0.585454818214395\\
3.25	0.561693316046611\\
3.3	0.541870193515853\\
3.35	0.526476009493944\\
3.4	0.515834396671455\\
3.45	0.510096314425544\\
3.5	0.509239047971736\\
3.55	0.513071008391787\\
3.6	0.521251234080092\\
3.65	0.533332976922581\\
3.7	0.548824694904986\\
3.75	0.567251376868286\\
3.8	0.588194136238961\\
3.85	0.611301103194983\\
3.9	0.636276679227066\\
3.95	0.662864562010084\\
4	0.690830796132134\\
};

\addplot [color=mycolor4, dashed, line width=2.0pt, forget plot]
  table[row sep=crcr]{%
0	1\\
0.05	0.997504688256135\\
0.1	0.990075051695429\\
0.15	0.977880318916352\\
0.2	0.961203614493663\\
0.25	0.940442225873442\\
0.3	0.916103021370078\\
0.35	0.888788317844103\\
0.4	0.859170501828846\\
0.45	0.827958428234553\\
0.5	0.795863706424717\\
0.55	0.763572767546511\\
0.6	0.731721988157524\\
0.65	0.700872263709862\\
0.7	0.671485988073257\\
0.75	0.643913003274498\\
0.8	0.618395376935233\\
0.85	0.595097476005087\\
0.9	0.574151022850046\\
0.95	0.555696771221645\\
1	0.539911905494496\\
1.05	0.527020179090475\\
1.1	0.517289838380014\\
1.15	0.511027277851244\\
1.2	0.508562764955894\\
1.25	0.510218397240443\\
1.3	0.516259534610572\\
1.35	0.526840773927428\\
1.4	0.541961926765607\\
1.45	0.561449183815197\\
1.5	0.584961694554891\\
1.55	0.612009787136765\\
1.6	0.64197647248475\\
1.65	0.674143023522835\\
1.7	0.707721551670726\\
1.75	0.741898274793052\\
1.8	0.775883431338136\\
1.85	0.808952429866362\\
1.9	0.840467141906507\\
1.95	0.869878749069559\\
2	0.89671987823239\\
2.05	0.920595769376319\\
2.1	0.941179530379951\\
2.15	0.958208429578104\\
2.2	0.971478479573238\\
2.25	0.980839039903705\\
2.3	0.986190151010834\\
2.35	0.987484657324858\\
2.4	0.984734354578939\\
2.45	0.978015713215658\\
2.5	0.967471338680122\\
2.55	0.953306535438179\\
2.6	0.935780854497847\\
2.65	0.915193419451449\\
2.7	0.891863984289644\\
2.75	0.866117029323415\\
2.8	0.83827643355237\\
2.85	0.808675451327283\\
2.9	0.777680027026996\\
2.95	0.74571370500819\\
3	0.713272299501338\\
3.05	0.680926085714755\\
3.1	0.649312355806215\\
3.15	0.619122334753549\\
3.2	0.591083103182584\\
3.25	0.565926378915289\\
3.3	0.544339331726615\\
3.35	0.526907717353464\\
3.4	0.514066860460254\\
3.45	0.506073346051701\\
3.5	0.503003759079056\\
3.55	0.504771447157579\\
3.6	0.511146686545277\\
3.65	0.521778586543338\\
3.7	0.536224923167955\\
3.75	0.553992399546518\\
3.8	0.574584826019227\\
3.85	0.597546724880505\\
3.9	0.622484843331295\\
3.95	0.64906421202376\\
4	0.67698953791247\\
};

\end{axis}
\end{tikzpicture}
    \caption{Solution for \(|a\pump(\xi)|^2\) to \Cref{eq:Eq_ap}, using \(\mu = 2\) and different number of harmonics included in computation, \(M = 2\) (blue), 3 (orange), 4 (yellow), 5 (dashed magenta).}
    \label{fig:PumpTransmissionSolutionsM}
\end{figure}

When the number of harmonics is restricted to \(M=2\), an analytical solution is available \cite{Bloembergen1962}, which shows that for a linear dispersion the injected tone is fully converted to the second harmonic on a length inversely proportional to the amplitude of the injected signal. In the presence of dispersion, both harmonics exhibit oscillatory behaviour (see blue curve in \Cref{fig:PumpTransmissionSolutionsM}) while preserving the quantity \(|A\pump(x)|^2 + 4|A\ph2(x)|^2 = \text{constant}\). Analytical solutions are also available for larger number of harmonics, but their behaviour becomes rather complex, so we resort to numerics. 

We perform a numerical study under the assumption that all relevant harmonics have frequencies well below the cutoff.
For a weak dispersion the pre-exponential factors in \Cref{eq:Eq_Ap} can be approximated with linear functions, 
\(k\ph m a = m\omega\pump/\omega_0\), while the exponential dephasing factors are approximated with the 
lowest order (cubic) corrections. The latter can be expressed through the dephasing 
\(\Delta\approx k\pump^3a^2/32\) defined in \Cref{eq:delta},
\begin{equation}
\begin{aligned}
    \Delta_{m,n} &= 8\Delta d_{m,n} = \frac{k\pump^3a^2}4 \, d_{m,n} \,,\\
    d_{m,n} &= \frac12 mn(m-n) \,. 
\end{aligned}
\end{equation}

To compute and analyse the solutions to \Cref{eq:Eq_Ap}, it is convenient to introduce dimensionless spatial coordinate and rescaled harmonic amplitudes,
\begin{equation}
    \xi = \frac{\varepsilon k\pump}4 x, \quad a\ph m(\xi(x)) = m \frac{A\ph m(x)}{A\pump(0)}\,,
\label{eq:x_xi}
\end{equation}
where \(\varepsilon\) is the pumping strength defined in \Cref{eq:epsilonAp}, which has the form in the low frequency limit, \(\varepsilon = |\chi_3 A\pump(0)|k\pump a\). Then \Cref{eq:Eq_Ap} reduces to a compact form,
\begin{equation}
\begin{aligned}
  (a_m)'_\xi &= m \sum_{n=m+1}^M a_n a^\ast_{n-m} \e^{\I\mu\xi d_{n,m}} \\
    &- \frac m2 \sum_{n=1}^{m-1} a_n a_{m-n} \e^{-\I\mu\xi d_{m,n}} \,,
\end{aligned}
\label{eq:Eq_ap}
\end{equation}
where the spatial behaviour of all harmonics is described with a \textit{single scaling parameter},
\begin{equation}
    \mu = \frac{32\Delta}{\varepsilon k\pump } \approx \frac{k\pump^2 a^2}\varepsilon\,.
\label{eq:mu}
\end{equation}
This parameter is proportional to the ratio of the dephasing and the nonlinear pumping strength, and has clear physical meaning indicating that it is the interplay between the dispersion and the nonlinear coupling (the pumping strength) that defines the behaviour of the harmonics. 

The differential equations in \Cref{eq:Eq_ap} are solved numerically for different values of \(\mu\) and \(M\) using 4th and 5th order Runge-Kutta methods of the MATLAB function \texttt{ode45}.

\vspace{2cm}
\begin{figure}[ht]
    \centering
    \begin{tikzpicture}\input{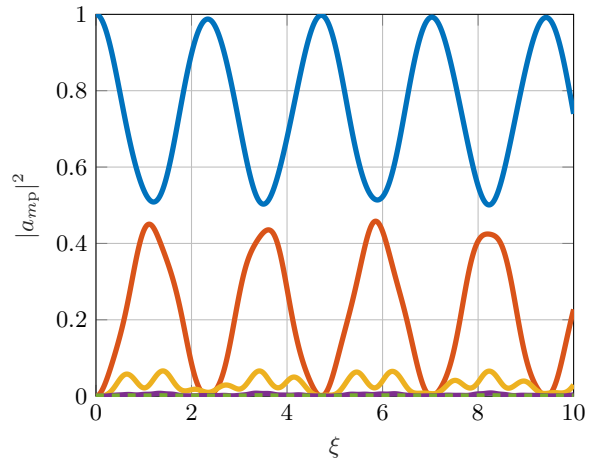}\end{tikzpicture}
    \caption{First five harmonics for \(\mu=2\) and \(M=5\): First (blue), second (orange), third (yellow), fourth (purple) and fifth (dashed green).}
    \label{fig:AllPumpHarmonicsTransmissionSolutions}
\end{figure}

The spatial dependence of the solutions truncated at \(M=5\) is shown in \Cref{fig:AllPumpHarmonicsTransmissionSolutions} for \(\mu=2\). All the harmonics oscillate but while amplitudes of the first and second harmonics are substantial, the amplitudes of higher harmonics, \(m=3,4,5\), are decreasingly small. Correspondingly, the effect of the latter on the main pump tone is small, as illustrated in \Cref{fig:PumpTransmissionSolutionsM} for \(\mu=2\) and different values of \(M\). Here the solution for the main pump tone coupled to three harmonics clearly differs from the one coupled to two harmonics, but the more harmonics are included the smaller effect they have on the solution for the main pump tone. The solution for the main pump tone appears to converge at \(M\gtrsim 5\). 

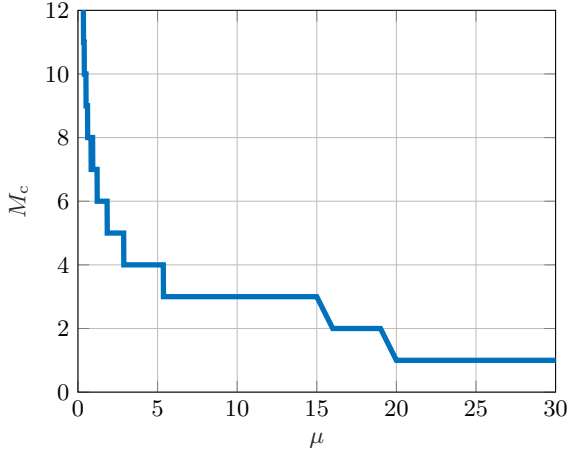
\begin{figure}[ht]
    \centering
    \begin{tikzpicture}
%
%
\definecolor{mycolor1}{rgb}{0.00000,0.44700,0.74100}%
%

\begin{axis}[%
width=2.5in,
height=2in,
at={(2.6in,1.103in)},
scale only axis,
xmin=0,
xmax=30,
xlabel style={font=\color{white!15!black}},
xlabel={\(\mu\)},
ymin=0,
ymax=12,
ylabel style={font=\color{white!15!black}},
ylabel={\(M_\mathrm{c}\)},
axis background/.style={fill=white},
xmajorgrids,
ymajorgrids
]
\addplot [color=mycolor1, line width=2.0pt, forget plot]
  table[row sep=crcr]{%
0.25	12\\
0.34	12\\
0.35	11\\
0.39	11\\
0.4	10\\
0.5	10\\
0.51	9\\
0.6	9\\
0.61	8\\
0.82	8\\
0.83	7\\
0.89	7\\
0.9	8\\
0.92	8\\
0.93	7\\
1.2	7\\
1.21	6\\
1.83	6\\
1.84	5\\
2.87	5\\
2.88	4\\
5.36	4\\
5.37	3\\
15	3\\
16	2\\
19	2\\
20	1\\
30	1\\
};
\end{axis}
\end{tikzpicture}
    \caption{The critical number of harmonics \(M\crit\) for different values of \(\mu\) for the length \(\xi=10\) and error tolerance \(0.01\). }
    \label{fig:Mcrit}
\end{figure}

\begin{figure}[ht]
    \centering
    \begin{tikzpicture}\input{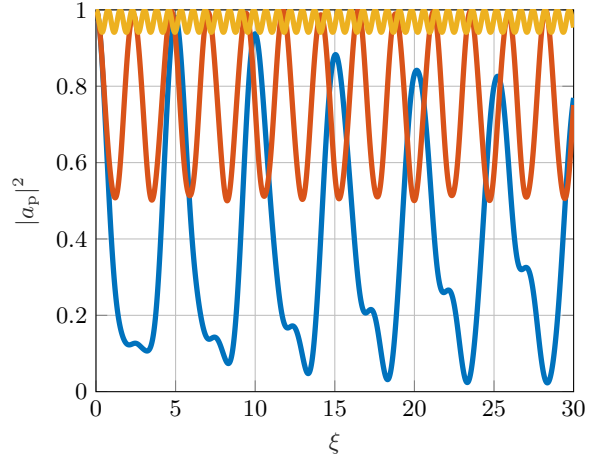}\end{tikzpicture}
    \caption{Solution for \(|a\pump(\xi)|^2\) to \Cref{eq:Eq_ap}, using \(\mu = 0.5\) (blue), 2 (orange), and 8 (yellow) and corresponding \(M\crit = 11,6,4\).}
    \label{fig:PumpTransmission}
\end{figure}

Our numerical studies show that the number of large-amplitude harmonics depends on the value of \(\mu\), the number of large-amplitude harmonics increases when \(\mu\) decreases, which corresponds to a weaker dispersion or stronger pumping. Furthermore, for a given \(\mu\), the amplitudes of harmonics with numbers exceeding a certain critical value, \(m>M\crit\), become negligible on a given length, as illustrated in \Cref{fig:Mcrit}. One can hence truncate \Cref{eq:Eq_ap} at \(M = M\crit(\mu)\) to accurately compute the solution for the main pump tone. The result of such a study is presented in \Cref{fig:PumpTransmission}. We solve \Cref{eq:Eq_ap} for certain values of \(\mu\) and find the corresponding values for \(M\crit\). For \(\mu = 0.5\) we find that \(M\crit = 10\), for \(\mu = 2\) we find that \(M\crit = 5\), and for \(\mu = 8\), \(M\crit = 3\). Different values of \(M\crit\) as a function of \(\mu\) is shown in \Cref{fig:Mcrit}.
As seen in \Cref{fig:PumpTransmission}, the pump behaviour can in general be summarised as follows: The pump oscillates between full transmission and some lower bound. The larger the \(\mu\), the smaller the oscillation amplitude and period.

%

\subsection{Comparison with four-wave mixing}
It is instructive to compare the pump harmonic generation studied in the previous section to the pump harmonic generation by the 4WM, where the phase mismatch introduced by the Kerr effect prevents exponential amplification. One should anticipate that a similar mechanism would suppress the generation of pump high harmonics \cite{Bloembergen1962}. We test this assumption by computing the third harmonic of the pump using the developed scaling method. 

\begin{figure}[ht]
    \centering
    \begin{tikzpicture}\input{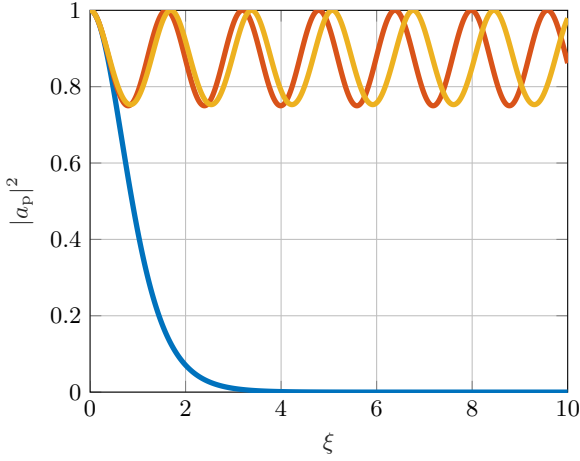}\end{tikzpicture}
    \caption{Spatial dependence of pump signal under 4WM including one up-converted mode for non-dispersive spectrum, \(\mu=0\) (yellow); for comparison the pump signal under 3WM is shown for \(\mu=0\) (blue), and for dispersive spectrum, \(\mu=3\) (orange).}
    \label{fig:Comparison2modes_3WMvs4WM}
\end{figure}

The equations for the pump harmonics are derived in a similar way as for the 3WM. Restricting to the third harmonic, we have, 
\begin{subequations}\label{eq:Ap4WM}
\begin{equation}
\begin{aligned}
    A\pump' &= \I\frac{\chi_4a^2}8 \left( k\pump^3 A\pump^2 A\pump^* 
    -  k\ph3 k\pump^2 A\ph3 A\pump^{*2} \e^{\I\left(k\ph3 - 3k\pump\right)x} \right) 
\end{aligned}
\end{equation}
\begin{equation}
\begin{aligned}
    A\ph3' &= \I\frac{\chi_4a^2}8 \left( 2k\ph3 k\pump^2 A\ph3 A\pump A\pump^* 
    - \frac13 k\pump^3 A\pump^3 \e^{-\I\left(k\ph3 - 3k\pump\right)x} \right) ,
\end{aligned}
\end{equation}
\end{subequations}
where \(\chi_4\) is the 4-th order nonlinearity coefficient derived in a similar way to \(\chi_3\) in \Cref{sec:equations}. The value of \(\chi_4\) is \(\frac12\) for a junction TWPA, while it is typically smaller for an rf-SQUID or a SNAIL TWPA at zero bias. With similar rescalings as in \Cref{sec:3WMPumpHarmonics}, 
\begin{equation}
    \xi = \frac{\chi_4A\pump(0)^2k\pump^3a^2}8 x, \quad a\ph m(\xi) = m \frac{A\ph m}{A\pump(0)} \,,
\end{equation}
we write these equations in a dimensionless form,
\begin{subequations}
\begin{align}
    (a\pump)'_\xi &= \I \left( a\pump^2a\pump^* - 3a\ph3a\pump^{*2} \e^{\I\mu\xi} \right), \\
    (a\ph3)'_\xi &= 3\I \left( 2a\ph3a\pump a\pump^* - \frac13 a\pump^3 \e^{-\I\mu\xi} \right)\,.
\end{align}
\end{subequations}
Here the spatial behaviour of both harmonics is defined by a single scaling parameter,
\begin{equation}
    \mu = \frac{256\Delta}{\chi_4A\pump(0)^2k\pump^3a^3} \approx \frac{8k\pump^2a^2}{\chi_4\theta\pump(0)^2}.
\end{equation}
We solve \Cref{eq:Ap4WM} numerically for different values of \(\mu\), the result is presented in \Cref{fig:Comparison2modes_3WMvs4WM}. In contrast to 3WM, where the first harmonic is completely depleted in the absence of frequency dispersion, \(\mu=0\) (blue line), it oscillates for the 4WM (yellow line), similar to 3WM at a considerable dispersion, \(\mu=3\) (orange line). 

Thus we conclude that the Kerr effect, caused by the \(\chi_4\)-term, not only prevents exponential amplification but also suppresses the generation of higher harmonics, which makes up-conversion processes less dangerous for 4WM amplification compared to 3WM.
%

\subsection{Full multimode model}

Now we proceed with the discussion of the amplification in the 3WM regime and assume a weak signal tone being injected in addition to the strong pump tone, \(A\signal(0) \ll A\pump(0)\). We perform our analysis under the same assumption as in \Cref{sec:Disc3modes} of small amplitudes of signal and idler compared to the pump amplitude within the whole TWPA, \(A\signal(x), A\idler(x) \ll A\pump(x)\). This assumption allows us to neglect the back-action of signal and idler on the pump (pump depletion), and also neglect the generation of signal and idler harmonics while allowing the up-conversion of signal and idler by the pump harmonics. The latter assumption implies a linearisation of the equations with respect to all signal and idler harmonics. 

With the adopted approximations, the field in the TWPA will consist of a linear combination of the pump and pump harmonics, \(m\omega\pump\), signal \(\omega\signal\), down-converted idler, \(\omega\idler = \omega\pump - \omega\signal\), and all their up-converted modes by the pump and pump harmonics, \(\omega\sh m =
\omega\signal + m\omega\pump\), \(\omega\ih m = \omega\idler + m\omega\pump\), see \Cref{fig:FreqDiagram}. The ansatz is therefore,
\begin{equation}
\begin{aligned}
    \phi(x,t) &= \frac12 \sum_{m=1}^M \left[ A\ph m \e^{\I k\ph mx - \I m\omega\pump t} + \text{c.c} \right] \\  
    &+ \frac12 \sum_{m=0}^{M-1} \left[ A\sh m \e^{\I k\sh mx - \I(\omega\signal+m\omega\pump)t} \right. \\
    &+ \left. A\ih m \e^{\I k\ih mx - \I(\omega\idler+m\omega\pump)t} + \text{c.c.} \right].
\end{aligned}
\end{equation}
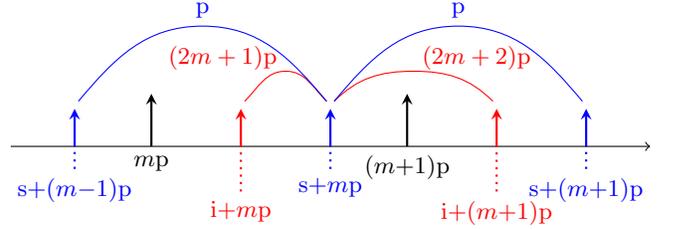
\begin{figure}[ht]
    \centering
    \begin{tikzpicture}
\renewcommand\freqscale{3.4}
\renewcommand\mixheightlower{0.6}
\renewcommand\mixheightupper{0.4}
\draw[->] (-4.25,0) -- (4.25,0);
\foreach\x in {0,1} {
    \draw[thick, -stealth] (\freqscale*\x-0.7*\freqscale,0) node[anchor=north]{
        \ifnum\x=0 {\(m\)p} \else{\((m\signed\x)\)p} \fi
    } -- +(0,0.7);
}
\foreach\x in {-1,0,1} {
    \draw[blue, dotted, thick] (\freqscale*\x,-0.3) node[anchor=north]{
        \ifnum\x=0 {s+\(m\)p} \else{s+\((m\signed\x)\)p} \fi
    } -- (\freqscale*\x,0);
    \draw[thick, blue, -stealth] (\freqscale*\x,0) -- +(0,0.5);
}
\foreach\x in {0,1} {
    \draw[red, dotted, thick] (\freqscale*\x-0.35*\freqscale,-0.6) node[anchor=north]{
        \ifnum\x=0 {i+\(m\)p} \else{i+\((m\signed\x)\)p} \fi
    } --  (\freqscale*\x-0.35*\freqscale,0);
    \draw[thick, red, -stealth] (\freqscale*\x-0.35*\freqscale,0) -- +(0,0.5);
}
\foreach\x in {0} {
    \draw[red] (\freqscale*\x-0.35*\freqscale+0.05,\mixheightlower) to[out=45,in=180] (\freqscale*\x-0.175*\freqscale,\mixheightlower+\mixheightupper) to[out=0,in=135] (\freqscale*\x-0.05,\mixheightlower);
    \draw[red] (\freqscale*\x-0.175*\freqscale,\mixheightlower+\mixheightupper-0.1) node[anchor=south east]{\((2m+1)\)p};
}
\foreach\x in {0} {
    \draw[red] (\freqscale*\x+0.05,\mixheightlower) to[out=45,in=180] (\freqscale*\x+0.325*\freqscale,\mixheightlower+\mixheightupper) to[out=0,in=135] (\freqscale*\x+0.65*\freqscale-0.05,\mixheightlower);
    \draw[red] (\freqscale*\x+0.325*\freqscale,\mixheightlower+\mixheightupper-0.1) node[anchor=south west]{\((2m+2)\)p};
}
\foreach\x in {0} {
    \draw[blue] (\freqscale*\x+0.05,\mixheightlower) to[out=45+5,in=180] (\freqscale*\x+0.5*\freqscale,\mixheightlower+2.5*\mixheightupper) node[anchor=south]{p} to[out=0,in=135-5] (\freqscale*\x+1*\freqscale-0.05,\mixheightlower);
    \draw[blue] (-\freqscale*\x-1*\freqscale+0.05,\mixheightlower) to[out=45+5,in=180] (\freqscale*\x-0.5*\freqscale,\mixheightlower+2.5*\mixheightupper) node[anchor=south]{p} to[out=0,in=135-5] (\freqscale*\x-0.05,\mixheightlower);
}\end{tikzpicture}
    \caption{Frequency diagram illustrating structure of \Cref{eq:Eq_As} for the signal \(m\)-th harmonic, \(a\sh m\); signal modes (blue), idler modes (red), pump harmonics (black), down-conversion processes (red) and up-conversion processes (blue). Each process is marked with the pump harmonic that is driving the process.}
    \label{fig:FreqDiagram}
\end{figure}

Using the reduced variables introduced in \Cref{eq:x_xi,eq:mu,eq:Delta} and detuning \(\delta\) defined in \Cref{eq:delta}, the equations for the signal modes take the form,
\begin{equation}
\begin{aligned}
    &(a\sh m)'_\xi = \left(m+\frac{1+\delta}2\right) \\
    \times &\left(\sum_{n=m+1}^M a\ph n a\ih{(n-m-1)}^* \e^{\I\mu\xi d_{n,m}^\text p}\right. \\
    &+ \left. \sum_{n=m+1}^{M-1} a\sh n a\ph{(n-m)}^* \e^{\I\mu\xi d_{n,m}^\text s} \right. \\
    &- \left. \sum_{n=1}^m a\ph n a\sh{(m-n)} \e^{-\I\mu\xi d_{m,(m-n)}^\text s} \right) \,,
\label{eq:Eq_As}
\end{aligned}
\end{equation}
where \(m \in [0, M-1]\). The numerical factors in the exponents are derived by using the cubic approximation of the dispersion relation and have the form,
\begin{equation}
\begin{aligned}
    &d_{n,m}^\text p = \frac{k\ph n - k\sh m - k\ih{(n-m-1)}}{8\Delta} \\
    &= \frac16 \left( n^3 - \left(m + \frac{1+\delta}2 \right)^3 - \left(n-m-1 + \frac{1-\delta}2 \right)^3 \right), \\
    & d_{n,m}^\text s = \frac{k\sh n - k\sh m - k\ph{(n-m)}}{8\Delta} \\
    &= \frac16 \left( \left( n + \frac{1+\delta}2 \right)^3 - \left( m + \frac{1+\delta}2 \right)^3 - (n-m)^3 \right) \,,
\end{aligned}
\end{equation}
where \(\Delta\) is defined in \Cref{eq:Delta}. The equations for the idler modes are similar, but with the replacements s\(\leftrightarrow\)i and \(\delta \leftrightarrow -\delta\). 

For \(M=1\), \Cref{eq:Eq_As} reduces to the continuous analog of the 3-mode model in \Cref{sec:Disc3modes},
\begin{equation}
\begin{aligned}
    (a\signal)'_\xi &= \frac{1+\delta}2 a\idler^\ast \e^{\I\mu\xi \left(1-\delta^2\right)/8} \\
    (a\idler)'_\xi &= \frac{1-\delta}2 a\signal^\ast \e^{\I\mu\xi \left(1-\delta^2\right)/8} \,.
\end{aligned}
\end{equation}
The solution has an exponential form, \(a\signal, \, a\idler \propto \e^{g\xi}\),  with the gain coefficient \cite{Tien1958},
\begin{equation}\label{eq:g3MM}
    g = \frac12 \sqrt{(1-\delta^2)\left[1 - \frac{\mu^2}{8^2} \left( 1 - \delta^2 \right)\right]} \,,
\end{equation}
that coincides with the long-wave length asymptotic in \Cref{eq:g_solution}. The exponential gain occurs only for small values of the scaling parameter given by, \(\mu < 8/\sqrt{1-\delta^2}\). 

\begin{figure}[ht]
    \centering
    \begin{tikzpicture}\input{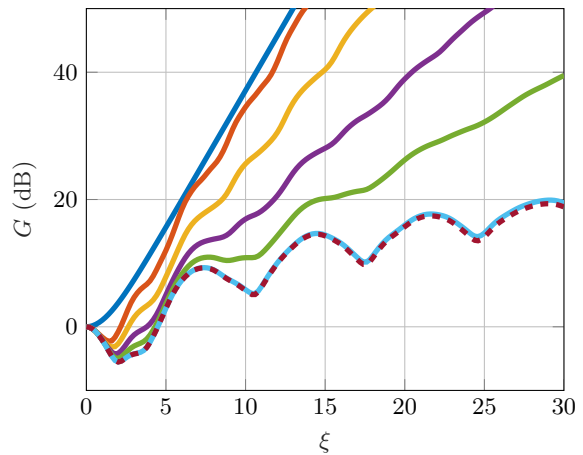}\end{tikzpicture}
    \caption{Gain \(G\) for \(\mu=1\) at zero detuning, \(\delta=0\), and \(M\) = 1, 4, 5, 6, 7, and 10 (from top to bottom), and \(M=11\) (dashed magenta).}
    \label{fig:QLGain_mu1}
\end{figure}

\begin{figure}[ht]
    \centering
    \begin{tikzpicture}\input{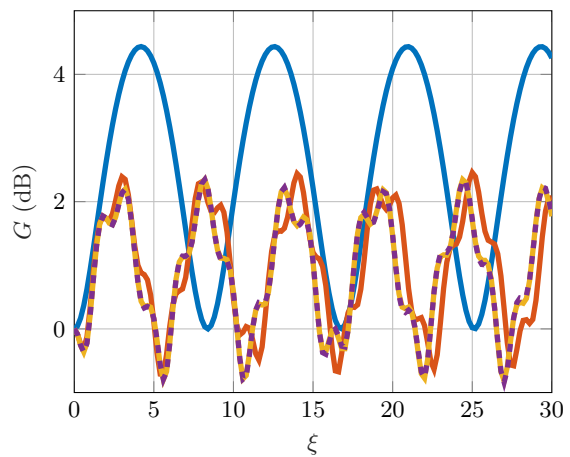}\end{tikzpicture}
    \caption{Gain \(G\) for \(\mu=10\) at zero detuning, \(\delta=0\), and \(M\) = 1 (blue), 2 (red), 3 (yellow), and 4 (dashed purple).}
    \label{fig:QLGain_mu10}
\end{figure}

We solve \Cref{eq:Eq_As} numerically in the same way as we solved them for the pump harmonics, including the spatial dependence of the pump harmonics. The results for the power gain as the function of length, \(G(\xi)=|a\signal(\xi)/a\signal[0]|^2\), are shown in \Cref{fig:QLGain_mu1,fig:QLGain_mu10} for \(\mu=1\) and \(\mu=10\) at zero detuning (\(\delta=0\)). 
These results demonstrate two distinctly different amplification regimes: For small values of \(\mu\), the gain grows on average with the TWPA length, but much slower than expected from 3-mode model, \Cref{eq:g3MM}. The gain suppression is the result of the up-conversion with many up-converted modes affecting the signal. In \Cref{fig:QLGain_mu1}, the gain reaches value \(G\sim20\,\text{dB}\) at scaled length \(\xi \approx 30\).
In the opposite case of large \(\mu\), the gain profile converges quickly, and the number of modes included in the simulation is small. Correspondingly, the gain spatial profile becomes oscillatory and has a relatively small amplitude, \Cref{fig:QLGain_mu10}. This reduction of the gain is the effect of phase mismatch, in accord with the criterion of a non-exponential amplification of the three-mode model, \(\mu>8\).

While one should not expect the gain larger than few dB for TWPA with any length when  \(\mu\) is large, for small \(\mu\) the gain grows with the length and interesting question is what gain can be achieved for realistic Josephson junction TWPA. The limitation is imposed by the necessity to accommodate all relevant modes within the spectral range. For the ten pump harmonics involved, as in \Cref{fig:QLGain_mu1}, the pump frequency should be limited, \(\omega\pump < 0.1\omega\crit\), which corresponds to \(ka < 0.2\). From \Cref{eq:x_xi,eq:mu}, we then deduce the real space length, \(x \sim 500a\xi = 15000\) unit cells. Such a long TWPA is unpractical. 

Summarising, we formulate the second important result of this paper: in weakly dispersive spectral region of low frequencies the gain above 20 dB is hard to achieve. For small \(\mu\), the gain is reduced by a strong up-conversion effect, while for large \(\mu\), the gain is small and oscillatory because of a large effective phase mismatch. 

%

\subsection{Comparison with experiment}
\label{sec:CompWithExperiment}
\begin{figure}[ht]
    \centering
    \begin{tikzpicture}\input{tikzpics/W34C19/v9_P-15.tikz}\end{tikzpicture}
    \caption{Measured data for a SNAIL-TWPA for the expected pump power \(P\pump^\text{exp} \approx -99\)\,dBm (black) and pump frequency \(\omega\pump/(2\pi) = 8.5\)\,GHz, and a theory fit (red) computed using the generalized quasi-linear model with \(\mu=15.18\) and \(\xi=2.77\), \Cref{eq:Eq_As,eq:Eq_ap}.}
    \label{fig:Fit_P-15}
\end{figure}
\begin{figure}[ht]
    \centering
    \begin{tikzpicture}\input{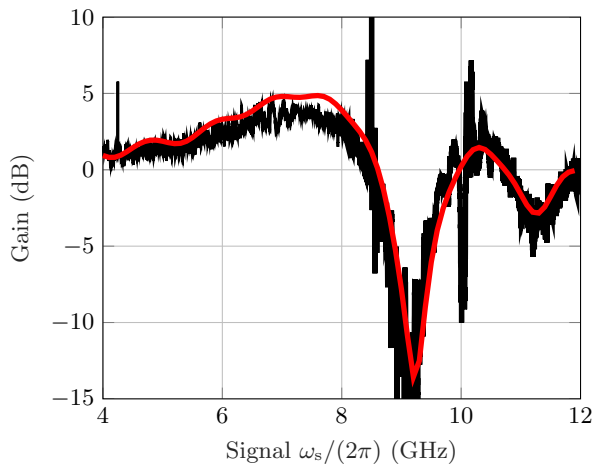}\end{tikzpicture}
    \caption{Measured data for a SNAIL-TWPA for the expected pump power \(P\pump^\text{exp} \approx -94\)\,dBm (black) and pump frequency \(\omega\pump/(2\pi) = 8.5\)\,GHz, and a theory fit (red) computed using the generalised quasi-linear model with \(\mu=8.54\) and \(\xi=4.93\), \Cref{eq:Eq_As,eq:Eq_ap}.}
    \label{fig:Fit_P-10}
\end{figure}
In this section we compare our theoretical predictions with experiments performed on a SNAIL-TWPA, \Cref{fig:TWPAs}d. The device consists of \(N=440\) unit cells, each unit cell contains one junction with \(I\crit[1] = 0.8\)\,\textmu A, and \(\mathcal N = 3\) identical junctions with \(I\crit[2] = 3\)\,\textmu A, \Cref{eq:LSn}. The biasing magnetic flux is applied at \(\Phi\approx0.4\Phi_0\), where \(\chi_3(\Phi) \approx 0.82\) (recall \Cref{eq:omega_Sn}), and the four-wave mixing vanishes. The pump frequency is placed at \(f\pump=8.5\)\,GHz, and the transmission (\(S_{21}\)) is measured using a VNA while sweeping the signal frequency within the band \(4-12\)\,GHz. We determine the gain by comparing the transmission of the signal for pump on versus pump off. The data is presented in \Cref{fig:Fit_P-15,fig:Fit_P-10} with black lines for the expected pump powers \(P\pump^\text{exp} \approx -99\)\,dBm and \(P\pump^\text{exp} \approx -94\)\,dBm, respectively. The red lines represent the theoretical fit.

The fitting is done by using \Cref{eq:Eq_As,eq:Eq_ap}. These equations contain two parameters, the scaling parameter \(\mu\), and the signal detuning \(\delta\). We generate the solution, \(a\signal(\xi; \delta, \mu)\) for chosen \(\delta\) and \(\mu\)  and compute the gain at the end of the chain for a range of detunings, \(G(\xi_\text{max},\delta,\mu)\). Then we sweep the two parameters, \(\xi_\text{max}\) and \(\mu\), to obtain the best fit to the data. Equivalently, one can use the parameters, \(k\pump a\) - normalised pump wave vector and 
\(\varepsilon\) - strength of coupling, which are related to \(\xi_\text{max}\) and \(\mu\) by virtue of \Cref{eq:x_xi} for the given length of the chain and \Cref{eq:mu}, in which we use the exact dispersion relation, \Cref{eq:dispersion_dc}, including the junction capacitances. 

The parameter values extracted from the fitting are presented in \Cref{tab:fitting}. The values of \(\varepsilon\) are found different for both datasets, which agrees quantitatively with the difference in the pump powers. At the same time, the value of \(k\pump a\) is the same in both cases as expected. The found value of the pump wave vector, \(k\pump a = 0.51\), however, differs from the theoretical value, \(k\pump a = 0.42\), computed from the dispersion relation, \Cref{eq:dispersion_dc} using the SNAIL parameters, \(\omega_S = 2\pi\cdot 20.6\)\,GHz, and \(C = 154\)\,fF and \(C\J = 17.9\,\)fF. We attribute this discrepancy to a non-uniform magnetic flux bias. The on-chip pump current, \(I\pump\), is determined by computing the pump-induced phase difference, \(|\theta\pump(0)| = \varepsilon/|\chi_3|\), \Cref{eq:epsilonAp}, and then connecting it to the current using \Cref{eq:Delta_Sn}. The found values of the on-chip pump power are consistent with the expected values, as the estimated loss of the line is \(\sim 84\)\,dB.

Summarising, the theory reproduces very well the measured frequency dependence of the gain, \(G(\omega\signal)\), despite that only two fitting parameters are at hand to describe an intricate interplay of the pump, signal, idler and their up-converted modes. Furthermore, our analysis reveals that the measurements are done in the regime of non-exponential amplification of large \(\mu\), \Cref{fig:QLGain_mu10}, which explains the small measured values of the gain. 

%
\begin{table}[H]
    \centering
    \caption{Parameters extracted from the fitting the data in \Cref{fig:Fit_P-15,fig:Fit_P-10}.}
    \label{tab:fitting}
    \begin{tabular}{|c|c|c|}
        \hline
        \(P\pump^\text{exp}\) (dBm) & \(-99\) & \(-94\) \\ \hline
        \(\varepsilon\) & 0.0504 & 0.0896 \\ \hline
        \(k\pump a\) & 0.51 & 0.51 \\ \hline
        \(\mu\) & 15.18 & 8.54 \\ \hline
        \(\xi_\text{max}\) & 2.77 & 4.93 \\ \hline
        \(I\pump\) (nA) & 52 & 93 \\ \hline
        \(P\pump\) (dBm) & \(-101.6\) & \(-96.6\) \\ \hline
        Fig. & \ref{fig:Fit_P-15} & \ref{fig:Fit_P-10} \\ \hline
    \end{tabular}
\end{table}
%

\section{A solution: two-band frequency spectrum}
\label{sec:solution}
In this section we revisit the 3-mode model of \Cref{sec:Disc3modes} and consider the possibility of reducing the dispersion at high frequencies, in the no-up-conversion region, \(\omega\pump > \Omega_\text{th} =4\omega_0/3\), to maintain the high gain. The idea is to create a sweet spot in the TWPA frequency spectrum where the signal injected at the degeneracy point, \(\omega\signal = \omega\pump/2\), is exactly phase matched with the pump, \(\kappa\signal = \kappa\pump/2\). Such a possibility does not exist for the TWPA studied in \Cref{sec:Disc3modes}. However, as we theoretically prove in this section, such a possibility appears for a TWPA with a two-band frequency spectrum (cf. \cite{GaoPRXQ2021}). At such a sweet spot a strong exponential amplification is predicted to occur and, moreover, it persists within a wide frequency band. 

A common way to create a gap in the TWPA spectrum is to periodically modulate the device parameters along the propagation direction. This is routinely done for kinetic inductance TWPAs by modulating the geometry, and thereby the impedance, of the transmission line \cite{LeDuc2012}. 
For the Josephson junction TWPAs another method is used – adding linear \(LC\)-oscillators to the TWPA cells. In this case, a spectral gap opens at the resonance frequency of the oscillators. This method of dispersion engineering is used in four-wave mixing devices to mitigate the Kerr effect \cite{Obrien2014,Macklin2015,Martinis2015}.

\begin{figure}[ht]
    \centering
    \begin{subfigure}{0.45\linewidth}
        \centering
        \begin{tikzpicture}    
\foreach\y in {1,-0.6} {\draw (0.6,\y) -- (3.5,\y);}
\draw[very thick,fill=white] (0.95,0.75) rectangle(1.45,1.25);
\draw (1.2,1) node{\(\mathcal L\J\)};
\draw (1.7,1) -- +(0,-1.6);
\capacitor{1.7}{0.55}
\ground{2.125}{-0.6}
\draw (2.8,1) -- (2.8,-0.6);
\draw[fill=white] (2.3,-0.4) rectangle(3.3,0.6);
\capacitor[\ensuremath{C_\mathrm{c}}]{2.8}{0.85}
\capacitor[\ensuremath{C_\mathrm{osc}}]{2.3}{0.15}
\inductor[\ensuremath{L_\mathrm{osc}}]{3.3}{0.5}
\foreach\x in {0.6,3.5} {\draw[fill] (\x,1) circle(0.05);}
\draw (0.6,1) node[anchor=south]{\(\phi_n\)};
\draw (3.5,1) node[anchor=south]{\(\phi_{n+1}\)};
\end{tikzpicture}
        \caption{}
        \label{fig:RPM}
    \end{subfigure}
    \begin{subfigure}{0.45\linewidth}
        \centering
        \begin{tikzpicture}
\def\nodeposdata{0.5 1.9 3.4}
\readarray\nodeposdata\nodepos[1,3]
\def\nodenamedata{-1 {} +1}
\readarray\nodenamedata\nodename[1,3]
\foreach\y in {1,-0.6} {\draw (0.5,\y) -- (3.5,\y);}
\ground{2}{-0.6}
\foreach\x in {1,2} {
    \draw[very thick,fill=white] (1.5*\x-0.6,0.7) rectangle(1.5*\x,1.3);
    \draw (1.5*\x-0.3,1) node{\(\mathcal L\J[\x]\)};
    \draw (1.5*\x+0.4,1) -- (1.5*\x+0.4,-0.6);
    \capacitor{1.5*\x+0.4}{0.3}
}
\foreach\x in {1,2,3} {
    \draw[fill] (\nodepos[1,\x],1) circle(0.05) node[anchor=south]{\(\phi_{n\nodename[1,\x]}\)};
}
\end{tikzpicture}
        \caption{}
        \label{fig:Mod}
    \end{subfigure}
    \caption{Circuit diagrams for TWPAs with two-band frequency spectrum; (a) \(LC\)-oscillators added to the unit cells, and (b) periodically modulated parameters of the unit cells.}
    \label{fig:TWPAvarations}
\end{figure}
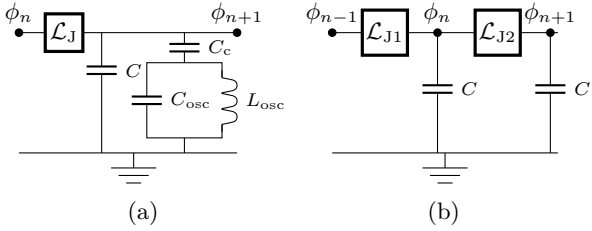
%

\subsection{Chain with oscillators}
\begin{figure}[ht]
    \centering
    \begin{tikzpicture}\input{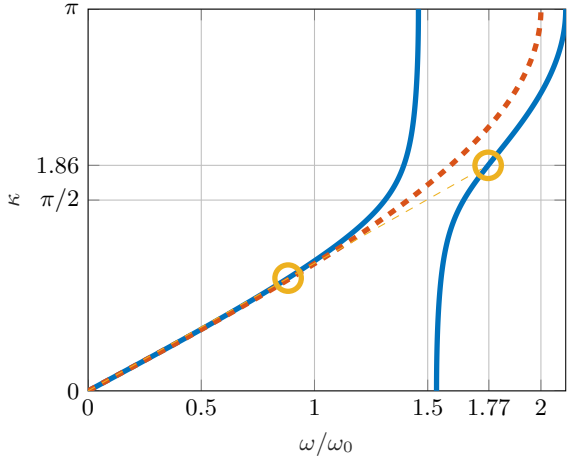}\end{tikzpicture}
    \caption{2-band frequency dispersion relation for a TWPA with \(LC\)-oscillators with \(\omega_1 = 1.5\omega_0\) and \(\nu = 0.95\) (solid lines). The dispersion relation exhibits a frequency gap at the resonance frequency of the \(LC\)-oscillator, the dashed line indicates the dispersion relation in the absence of oscillators, the circles indicate the positions of phase matched points which lie on a straight line within the no-up-conversion region (shown in \Cref{fig:UpconversionDisabling}).}
    \label{fig:spectrumRPM}
\end{figure}

Let us consider the TWPA with \(LC\)-oscillators. The corresponding circuit is presented in \Cref{fig:RPM}. The derivation of the dynamical equation is straightforward: we add the oscillator circuit variables to the Lagrangian, \Cref{eq:general_L}, and eliminate the oscillator variables from the dynamic equations. Specifically we consider the dc current biased TWPA. In this case, \Cref{eq:As_eqn} for the chain without oscillators remains valid with the only difference, the factor \(\omega^2/\omega_0^2\) in the first term is replaced with
\begin{equation}
    \frac{\omega^2}{\omega_0^2} \to \frac{\omega^2}{\omega_0^2} \frac{\nu\omega^2 - \omega_1^2}{\omega^2- \omega_1^2}
\end{equation}
where
\begin{align}
    \omega_1^2 &= \frac1{L_\text{osc}(C_\text{osc}+C_\text{c})}, \\
    \nu &= 1 - \frac{C_\text{c}^2}{(C+C_\text{c})(C_\text{osc}+C_\text{c})} \,. 
\end{align}
The dispersion equation then takes the form
\begin{equation}
    4\omega_0^2\sin^2\frac\kappa2 = \omega^2 \frac{\nu\omega^2- \omega_1^2}{\omega^2- \omega_1^2}\,.
\label{eq:dispersion}
\end{equation}
Solving for \(\omega\) we get
\begin{equation}
\begin{aligned}
    \omega^2_\pm &= \frac1{2\nu} \left[ \left(\omega_1^2 + 4\omega_0^2\sin^2 \frac\kappa2 \right) \right. \\
    &\pm \left. \sqrt{\left(\omega_1^2 + 4\omega_0^2\sin^2\frac\kappa2 \right)^2 - 16\nu\omega_1^2\omega_0^2\sin^2 \frac\kappa2} \,\right] \,.
\label{eq:spectrumRPM}
\end{aligned}
\end{equation}
The derived spectrum is depicted in \Cref{fig:spectrumRPM}; it consists of two bands separated by a gap. 

To identify the sweet spot we assume the pump frequency within the upper band, and the signal frequency within the lower band, and solve equation, \(\omega_-(\kappa\pump/2) = \omega_+(\kappa\pump)/2 \).
Converted for the pump frequency, this equation has the explicit form
\begin{equation}
    \frac{3(1- \nu)}{\omega\pump^2 - \omega_1^2} = \frac1{16\omega_0^2\omega_1^2} \frac{(4\omega_1^2 - \nu\omega\pump^2)^2}{4\omega_1^2 - \omega\pump^2} \,.  
\end{equation}
One can check by direct calculation that the solution indeed possesses the property, \(\kappa\signal = \kappa\pump/2\), as illustrated in \Cref{fig:spectrumRPM}: the pump and signal points are located on a straight line. Moreover, the pump frequency is located in the no-up-conversion frequency region (recall \Cref{fig:UpconversionDisabling}).
\begin{figure}[ht]
    \centering
    \begin{tikzpicture}\input{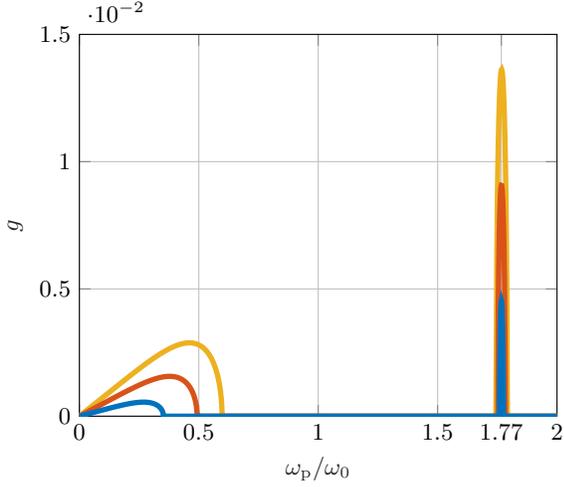}\end{tikzpicture}
    \caption{Gain coefficient for a signal at zero detuning, \(\delta=0\), for a TWPA with a two-band dispersion relation with \(\omega_1=1.5\omega_0\), \(\nu = 0.95\) and \(\varepsilon=0.02\) (blue), 0.04 (orange) and 0.06 (yellow). When the pump is placed within the lower band, the gain coefficient is similar to the case in the absence of oscillators (curves at the lower left corner). The sharp high-amplification peak emerges when the pump is placed within the upper band at the sweet spot, \(\omega\pump = 1.77\omega_0\). The peak widths are \(0.018\omega_0\), \(0.032\omega_0\) and \(0.048\omega_0\) respectively.}
    \label{fig:IncrementRPM}
\end{figure}
\begin{figure}[ht]
    \centering
    \begin{tikzpicture}
%
%
\definecolor{mycolor4}{rgb}{0.00000,0.44700,0.74100}%
\definecolor{mycolor3}{rgb}{0.85000,0.32500,0.09800}%
\definecolor{mycolor2}{rgb}{0.92900,0.69400,0.12500}%
\definecolor{mycolor1}{rgb}{0.49400,0.18400,0.55600}%
%

\begin{axis}[%
width=2.5in,
height=2in,
at={(2.6in,1.103in)},
scale only axis,
xmin=-1,
xmax=1,
xlabel style={font=\color{white!15!black}},
xlabel={\(\delta\)},
ymin=0,
ymax=0.015,
ylabel style={font=\color{white!15!black}},
ylabel={\(g\)},
axis background/.style={fill=white},
xmajorgrids,
ymajorgrids,
legend style={at={(0.03,0.97)}, anchor=north west, legend cell align=left, align=left, draw=white!15!black}
]
\addplot [color=mycolor2, line width=2.0pt, forget plot]
  table[row sep=crcr]{%
-1	0\\
-0.232	0\\
-0.224	0.00336637\\
-0.216	0.00596218\\
-0.208	0.00754647\\
-0.2	0.00870658\\
-0.192	0.00961036\\
-0.184	0.01033627\\
-0.176	0.01092934\\
-0.168	0.01141865\\
-0.16	0.01182452\\
-0.152	0.01216194\\
-0.144	0.01244246\\
-0.136	0.01267521\\
-0.128	0.01286764\\
-0.12	0.01302592\\
-0.112	0.01315527\\
-0.104	0.01326012\\
-0.1	0.01330458\\
-0.096	0.0133443\\
-0.088	0.01341112\\
-0.08	0.01346348\\
-0.072	0.01350388\\
-0.064	0.01353453\\
-0.056	0.01355732\\
-0.048	0.01357388\\
-0.04	0.01358561\\
-0.032	0.01359366\\
-0.024	0.01359896\\
-0.016	0.01360224\\
-0.008	0.01360401\\
0	0.01360456\\
0.008	0.01360401\\
0.016	0.01360224\\
0.024	0.01359896\\
0.032	0.01359366\\
0.04	0.01358561\\
0.048	0.01357388\\
0.056	0.01355732\\
0.064	0.01353453\\
0.072	0.01350388\\
0.08	0.01346348\\
0.088	0.01341112\\
0.096	0.0133443\\
0.1	0.01330458\\
0.104	0.01326012\\
0.112	0.01315527\\
0.12	0.01302592\\
0.128	0.01286764\\
0.136	0.01267521\\
0.144	0.01244246\\
0.152	0.01216194\\
0.16	0.01182452\\
0.168	0.01141865\\
0.176	0.01092934\\
0.184	0.01033627\\
0.192	0.00961036\\
0.2	0.00870658\\
0.208	0.00754647\\
0.216	0.00596218\\
0.224	0.00336637\\
0.232	0\\
1	0\\
};

\addplot [color=mycolor3, line width=2.0pt, forget plot]
  table[row sep=crcr]{%
-1	0\\
-0.192	0\\
-0.184	0.00287974\\
-0.176	0.00453071\\
-0.168	0.00557709\\
-0.16	0.00634017\\
-0.152	0.00692603\\
-0.144	0.00738703\\
-0.136	0.00775427\\
-0.128	0.00804834\\
-0.12	0.00828393\\
-0.112	0.00847208\\
-0.104	0.00862143\\
-0.1	0.00868377\\
-0.096	0.00873891\\
-0.088	0.00883023\\
-0.08	0.00890016\\
-0.072	0.00895274\\
-0.064	0.00899142\\
-0.056	0.00901912\\
-0.048	0.00903832\\
-0.04	0.00905115\\
-0.032	0.00905932\\
-0.024	0.00906424\\
-0.016	0.009067\\
-0.008	0.00906835\\
0	0.00906875\\
0.008	0.00906835\\
0.016	0.009067\\
0.024	0.00906424\\
0.032	0.00905932\\
0.04	0.00905115\\
0.048	0.00903832\\
0.056	0.00901912\\
0.064	0.00899142\\
0.072	0.00895274\\
0.08	0.00890016\\
0.088	0.00883023\\
0.096	0.00873891\\
0.1	0.00868377\\
0.104	0.00862143\\
0.112	0.00847208\\
0.12	0.00828393\\
0.128	0.00804834\\
0.136	0.00775427\\
0.144	0.00738703\\
0.152	0.00692603\\
0.16	0.00634017\\
0.168	0.00557709\\
0.176	0.00453071\\
0.184	0.00287974\\
0.192	0\\
1	0\\
};

\addplot [color=mycolor4, line width=2.0pt, forget plot]
  table[row sep=crcr]{%
-1	0\\
-0.136	0\\
-0.128	0.00207711\\
-0.12	0.00283051\\
-0.112	0.00332032\\
-0.104	0.00366707\\
-0.1	0.0038033\\
-0.096	0.00392008\\
-0.088	0.00410627\\
-0.08	0.00424279\\
-0.072	0.00434163\\
-0.064	0.0044117\\
-0.056	0.00445992\\
-0.048	0.0044918\\
-0.04	0.0045118\\
-0.032	0.00452348\\
-0.024	0.0045297\\
-0.016	0.00453262\\
-0.008	0.00453376\\
0	0.00453404\\
0.008	0.00453376\\
0.016	0.00453262\\
0.024	0.0045297\\
0.032	0.00452348\\
0.04	0.0045118\\
0.048	0.0044918\\
0.056	0.00445992\\
0.064	0.0044117\\
0.072	0.00434163\\
0.08	0.00424279\\
0.088	0.00410627\\
0.096	0.00392008\\
0.1	0.0038033\\
0.104	0.00366707\\
0.112	0.00332032\\
0.12	0.00283051\\
0.128	0.00207711\\
0.136	0\\
1	0\\
};

\end{axis}
\end{tikzpicture}
    \caption{Gain coefficient as function of detuning for a TWPA with two-band dispersion relation with \(\omega_1=1.5\omega_0\) and \(\nu = 0.95\), for pump at optimal phase matching point \(\approx1.77\omega_0\) and different pumping strengths, \(\varepsilon=0.02\) (blue), 0.04 (orange), and 0.06 (yellow). The corresponding bandwidths are \(\omega\signal-\omega\pump/2 = (0.14,\, 0.19, \, 0.23)\,\omega\pump\), respectively.}  
    \label{fig:BandwidthRPM}
\end{figure}
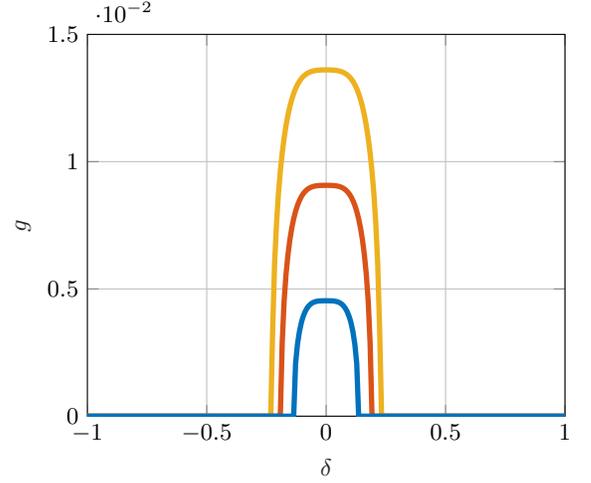

The equations for the gain coefficient, \Cref{eq:g_solution,eq:g_degen}, derived in \Cref{sec:Disc3modes} do not change their form in the present case; however, the dependence of the gain coefficient on the frequency is now different due to the different dispersion relation. Dependence of the gain coefficient at the signal degeneracy on the pump frequency is illustrated in \Cref{fig:IncrementRPM}. As soon as the pump is placed within the lower frequency band, the gain coefficient behaves similarly to the one in \Cref{fig:IncrementVsPumpFreq_Ap} for a TWPA without oscillators. However, when the pump is placed within the upper band at the sweet spot, the amplification dramatically increases up to \(g\approx 0.014\)
for a rather weak pumping strength, \(\varepsilon = 0.06\), about seven times smaller than the maximum pumping strength in \Cref{fig:IncrementVsPumpFreq_Ap} and definitely achievable in experiments. This \(g\)-value corresponds to the gain \(G\approx 24\)\,dB for \(N = 200\)\,unit cells. The pump position is rather flexible, the gain remains high within the pump bandwidth \(\sim 0.05\omega_0\).

For a detuned signal, large amplification persists within a quite wide frequency band, \(\sim 0.5 \omega\pump\) that could be of order of few GHz, as shown in \Cref{fig:BandwidthRPM}. This is due to a relatively weak dispersion in the low frequency region. 

\subsection{Sweet spot in periodically modulated chain}
Here we examine a periodically modulated TWPA and prove that there also exists a sweet spot. Experimental realisation of such a device will be reported elsewhere \cite{AnitasArticle}. The circuit is presented in \Cref{fig:Mod}, here each unit cell consists of two subcells with different Josephson junction parameters. Consider the dc current biased TWPA with different Josephson inductances in the subcells. The Lagrangian can be written for odd and even circuit nodes in analogy with \Cref{eq:general_L,eq:JJ_L},
\begin{eqnarray}
\begin{aligned}
    \mathcal L\J{} [\theta_n] &= \sum_{2n} \frac1{L\J[1]} \cos(\theta_{01} +\theta_{2n}) \\
    &+ \sum_{2n+1} \frac1{L\J[2]} \cos(\theta_{02} + \theta_{2n+1}) \,.
\end{aligned}
\end{eqnarray}
The biasing phases here obey the equations, \(\sin\theta_{0j} = I\dc/I_{\mathrm{c}j}\). Dynamical equations for such a chain have the form,
\begin{equation}
\begin{aligned}
    &-\ddot\phi_{2n} - \left( \omega_1^2\theta_{2n} - \omega_2^2\theta_{2n+1} \right) \\
    &+ \frac12 \left( \omega_1^2 \tan\theta_{01} \theta_{2n}^2 - \omega_2^2 \tan\theta_{02}\theta_{2n+1}^2 \right) = 0\,,
\end{aligned}
\label{eq:2band_eq}
\end{equation}
\begin{equation}
\begin{aligned}
    & -\ddot\phi_{2n+1} - \left( \omega_2^2 \theta_{2n+1} - \omega_1^2\theta_{2n+2} \right) \\
    &+ \frac12 \left( \omega_2^2 \tan\theta_{02} \theta_{2n+1}^2 - \omega_1^2 \tan\theta_{01} \theta_{2n+2}^2 \right) = 0 \,,
\end{aligned}
\label{eq:2band_eq2}
\end{equation}
where \(\omega_j^2 = \cos\theta_{0j}/ L\J[j]C\) are the subcell resonance frequencies.
The linearised equations define the spectral properties of the chain. Assume the solution having the form, \(\phi_{2n} = A\e^{\I \kappa (2n) - \I\omega t}\), \(\phi_{2n+1} = B\e^{\I \kappa (2n+1) - \I\omega t}\), then equations for amplitudes, \(A\) and \(B\), read,
\begin{equation}
\begin{aligned}
    \left( \omega^2 - \omega_1^2 - \omega_2^2)A + (\omega_1^2 \e^{-\I \kappa} + \omega_2^2 \e^{\I \kappa} \right) B &= 0 \\
    \left(\omega^2 - \omega_2^2 - \omega_1^2) B + (\omega_2^2 \e^{-\I \kappa} + \omega_1^2 \e^{\I \kappa} \right) A &= 0 \,,
\end{aligned}
\label{eq:ab_eq}
\end{equation}
and the dispersion relation has the form,
\begin{equation}
    \omega^2 = \left( \omega_1^2 + \omega_2^2 \right) \pm \sqrt{\left( \omega_1^2 + \omega_2^2 \right)^2 - 4\omega_1^2 \omega_2^2\sin^2 \kappa} \,. 
\label{eq:spectrum2band}
\end{equation}
\begin{figure}[ht]
    \centering
    \begin{tikzpicture}\input{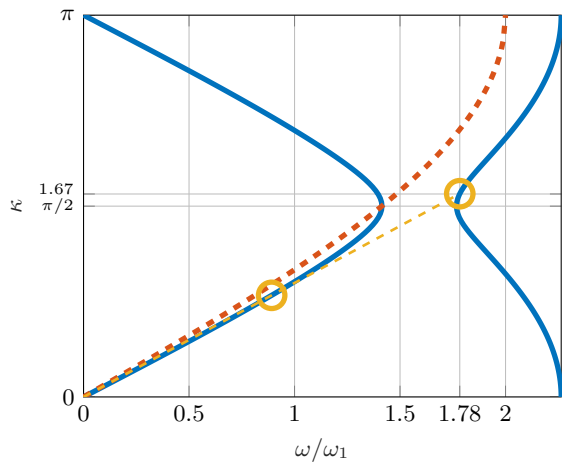}\end{tikzpicture}
    \caption{Dispersion relation of a periodically modulated Josephson junction chain with \(\omega_2 = 1.25\omega_1\); it consists of two bands separated by a gap (solid lines), dashed line indicates the dispersion relation in the absence of modulation, the circles indicate positions of phase matched points, they lie on a straight line within the no-up-conversion frequency region.}
    \label{fig:spectrumMod}
\end{figure}

The dispersion relation is depicted in \Cref{fig:spectrumMod}; it has two bands separated by a gap. The sweet spot is found from equations, \(\omega\signal = \omega\pump/2\), \(\kappa\signal = \kappa\pump/2\), whose solution reads, 
\begin{align}
    \omega\pump^2 &= 8\left[ \left( \omega_1^2 + \omega_2^2 \right) - \sqrt3 \omega_1\omega_2 \right] \label{eq:matching_pump} \,, \\
    \sin^2\frac{\kappa\pump}2 &= \frac{\sqrt3 \left[ \left(\omega_1^2 + \omega_2^2 \right) - \sqrt3 \omega_1\omega_2 \right]}{\omega_1 \omega_2}\,. \label{eq:matching_k}
\end{align}
This solution is illustrated in \Cref{fig:spectrumMod}: the positions of the pump and the signal at the degeneracy are indicated with the circles; they lie on a straight line at different sides of the gap in the region where the up-conversion is not possible. The gain in this setup is qualitatively similar to the TWPA with \(LC\)-oscillators.



\section{Conclusion}
\label{sec:conclusion}
In this paper, we propose a method for achieving a high gain for a lumped-element TWPA operating in the 3WM regime. The simple model of amplification in weakly dispersive medium \cite{Suhl1958,Tien1958}, relevant for kinetic inductance TWPAs and Josephson junction TWPAs at low frequencies predicts the gain up to 40 dB for a chain with 100 unit cells and a reasonable pump intensity. However, in practice, such a gain was never demonstrated. This was explained with a parasitic effect of generation of high harmonics and up-conversion processes \cite{Dixon2020}. We performed a detailed theoretical analysis of this regime including multiple pump harmonics and signal and idler up-converted modes. We identified a scaling parameter \(\mu\) that controls the gain and quantifies an interplay between the dispersion and the nonlinear wave interaction and found that the gain is strongly reduced for both small values as well as large values of \(\mu\), although for different reasons. When the dispersion is weak in relation to the interaction, \ie\ for small \(\mu\), the generation of up-converted modes is prominent. In the opposite limit of strong dispersion in relation to the interaction, \ie\ for large \(\mu\), the phase mismatch becomes the dominant effect. This finding is supported by the experimental observations on a SNAIL-TWPA, and the data are in quantitative agreement with the theoretical simulations. 

Our proposal concerns a different operation regime for which the cutoff frequency of the TWPA plays the central role. We proposed to place the pump close to the cutoff such that generation of up-converted modes is inhibited. Then, by solving the difference equations for discrete Josephson junction chain we found that there is a sweet spot where the pump and the signal are exactly phase matched. The sweet spot was proven to exist when the TWPA frequency spectrum consists of two bands separated by a gap. Studying different ways of engineering the two-band spectrum - by adding \(LC\)-oscillators or periodically modulating the chain parameters, we predicted that the gain at the sweet spot may achieve the values of order 25 dB within a few GHz amplification bandwidth for a chain with \(\sim 200\)\,unit cells and for moderate pump intensities. 
 
\section{Acknowledgements}
The project was supported by the Knut and Alice Wallenberg foundation via the Wallenberg Centre for Quantum Technology, and the EU Consortium OpenSuperQ. The authors acknowledge the use of the Nanofabrication Laboratory (NFL) at Chalmers University of Technology.


\end{document}